\documentclass[twocolumn,superscriptaddress,preprintnumbers,nofootinbib,amsmath,amssymb,aps,prmaterials,floatfix]{revtex4-2}

\usepackage{booktabs,siunitx}
\usepackage{subfig}
\usepackage{dcolumn} 
\usepackage{float}
\usepackage{comment}
\usepackage{ragged2e}
\usepackage{multirow}

\usepackage[table]{xcolor}

\usepackage{epsfig}
\usepackage{fixltx2e}

\usepackage{lmodern} 
\usepackage[T1]{fontenc}
\usepackage[utf8]{inputenc}

\usepackage[english]{babel}

\usepackage{xfrac}
\usepackage[version=4]{mhchem}
\usepackage{chemformula}

\usepackage[final=true,
					bookmarks=true,
					pdftoolbar=true,
					pdfmenubar=true,
					pdftitle={My title},
					pdffitwindow=true,
					pdfauthor={Author},
					colorlinks=true,
					linkcolor=blue,
					citecolor=blue,
					filecolor=blue,
					urlcolor=blue]{hyperref}
\usepackage{cleveref}

\usepackage{todonotes}

\usepackage{placeins}
\usepackage{amsmath,amssymb,amsfonts}
\usepackage{algorithmic}
\usepackage{graphicx}
\usepackage{textcomp}
\usepackage[table]{xcolor}
\usepackage{booktabs}
\usepackage[version=4]{mhchem}
\usepackage{tabularx,booktabs}
\usepackage{adjustbox}
\usepackage{gensymb}
\usepackage{tablefootnote}

\begin{document}

\title{Ge-based clinopyroxene series: first principles and experimental local probe study}

\author{Ricardo P. Moreira}
\affiliation{IFIMUP, Institute of Physics for Advanced Materials, Nanotechnology and Photonics, Departamento
de Física e Astronomia da Faculdade de Ciências da Universidade do Porto, Rua do Campo Alegre s/n,
4169-007 Porto, Portugal.}

\author{E. Lora da Silva}
\affiliation{IFIMUP, Institute of Physics for Advanced Materials, Nanotechnology and Photonics, Departamento
de Física e Astronomia da Faculdade de Ciências da Universidade do Porto, Rua do Campo Alegre s/n,
4169-007 Porto, Portugal.}
\affiliation{High Performance Computing Chair, University of Évora, Rua Romão Ramalho 59, 7000-671 Évora, Portugal}

\author{Gonçalo N. P. Oliveira}
\affiliation{IFIMUP, Institute of Physics for Advanced Materials, Nanotechnology and Photonics, Departamento
de Física e Astronomia da Faculdade de Ciências da Universidade do Porto, Rua do Campo Alegre s/n,
4169-007 Porto, Portugal.}

\author{P. Neenu Lekshmi}
\affiliation{IFIMUP, Institute of Physics for Advanced Materials, Nanotechnology and Photonics, Departamento
de Física e Astronomia da Faculdade de Ciências da Universidade do Porto, Rua do Campo Alegre s/n,
4169-007 Porto, Portugal.}

\author{Pedro~Rocha-Rodrigues}
\affiliation{IFIMUP, Institute of Physics for Advanced Materials, Nanotechnology and Photonics, Departamento
de Física e Astronomia da Faculdade de Ciências da Universidade do Porto, Rua do Campo Alegre s/n,
4169-007 Porto, Portugal.}

\author{Fábio G. Figueiras}
\affiliation{IFIMUP, Institute of Physics for Advanced Materials, Nanotechnology and Photonics, Departamento
de Física e Astronomia da Faculdade de Ciências da Universidade do Porto, Rua do Campo Alegre s/n,
4169-007 Porto, Portugal.}

\author{Abderrazzak Ait Bassou} \affiliation{CQ-VR Centro de Química Vila Real, School of Science and Technology (ECT), Physics Department, Universidade de Trás-os-Montes e Alto Douro, 5000-801 Vila Real, Portugal}

\author{Alessandro Stroppa}
\affiliation{CNR-SPIN c/o Università degli Studi dell’Aquila, Via Vetoio 10, 67010 Coppito, L’Aquila, Italy.}

\author{Claire V. Colin}
\affiliation{Université Grenoble Alpes, CNRS, Institut Néel, 38000, Grenoble, France}

\author{Céline Darie}
\affiliation{Université Grenoble Alpes, CNRS, Institut Néel, 38000, Grenoble, France}

\author{João G. Correia}
\affiliation{C2TN, DECN, Instituto Superior Técnico, Universidade de Lisboa, Bobadela, Portugal.}

\author{Lucy~V.~C.~Assali}

\affiliation{Instituto de Física, Universidade de São Paulo, CP 66318, 05315-970, São Paulo-SP, Brazil}

\author{Helena M. Petrilli}
\affiliation{Instituto de Física, Universidade de São Paulo, CP 66318, 05315-970, São Paulo-SP, Brazil}

\author{Armandina M. L. Lopes}
\affiliation{IFIMUP, Institute of Physics for Advanced Materials, Nanotechnology and Photonics, Departamento
de Física e Astronomia da Faculdade de Ciências da Universidade do Porto, Rua do Campo Alegre s/n,
4169-007 Porto, Portugal.}

\author{João P. Araújo}
\email{jearaujo@fc.up.pt}
\affiliation{IFIMUP, Institute of Physics for Advanced Materials, Nanotechnology and Photonics, Departamento
de Física e Astronomia da Faculdade de Ciências da Universidade do Porto, Rua do Campo Alegre s/n,
4169-007 Porto, Portugal.}

\date{\today}

\begin{abstract}
The electronic properties of the Ca/Sr and Mn site substitution of \ce{CaMnGe2O6}  and \ce{SrMnGe2O6} clinopyroxene systems have been investigated by \textit{ab-initio} calculations within the density functional theory (DFT) framework, using on-site Hubbard \textit{U} to describe the highly correlated Mn $3d$-states and a hybrid exchange-correlation functional to obtain the energy band gap values. Compositions such as \ce{(Ca,Sr)_{1-x}Cd_{x}MnGe2O6} and \ce{(Ca, Sr)Mn_{1-x}Cd_xGe2O6} (where $x=0.125, 0.25$) are predicted to be stable. Also, we proved that implanted Cd impurity could indeed replace either the Ca/Sr or the Mn sites in the crystalline structures. These findings were obtained by combining first principles electric field gradient calculations, using a supercell scheme, with experimental TDPAC results. Additionally DFT calculations showed that Cd substitution is expected to lead to a reduction in the band gap width. For the first time, the Cd-doped systems were successfully synthesized and experimental results evidencing opportunities for potential band-gap engineering are reported.

\end{abstract}


\maketitle

\section{Introduction}
\label{Sec:Intro}

Pyroxenes are a class of materials with general formula \ce{\textit{AMX}2O6},  where \textit{A} is a monovalent or a divalent cation, \textit{M} is respectively a trivalent or a divalent cation, and \textit{X} is typically either Si or Ge. They are well known in mineralogy and geology as they are one of the main rock-forming minerals of the Earth's crust~\cite{Streltsov2008}. Within this rich family of compounds, monoclinic pyroxenes (clinopyroxenes), with a $3d$ transition metal at the \textit{M}-site, have recently been the subject of interest in the field of condensed matter physics due to the diversity of their magnetic properties, 
namely: multiferroicity in \ce{SrMnGe2O6}~\cite{Ding_JMCC2016,PhysRevB.101.235109} and \ce{NaFe(Si/Ge)2O6}~\cite{Jodlauk2007,Kim2012}, magnetoelectric effect in \ce{CaMnGe2O6}~\cite{Ding_PRB2016} and \ce{Li(Cr/Fe)Si2O6}~\cite{Jodlauk2007}, and 
ferrotoroidal ordering in \ce{LiFeSi2O6}. 

In spite of their natural prevalence, pyroxenes have not been the subject of much computational or theoretical investigations. It is noteworthy of mentioning, that 
density functional theory (DFT) studies \cite{https://doi.org/10.1002/cphc.201701155} have shown that substitution of \ce{Al\textsuperscript{3+}} by \ce{Tl^{3+}} in \ce{NaAlSi2O6}  results in a reduction in the band gap width from 5.32 eV to 2.05 eV. A further reduction is observed when replacing \ce{Na} by the organic cation \ce{CH3SH2}, showing the high tunability of the band gap in these materials. Other works have shown their potential usefulness in batteries, demonstrating that pyroxenes \ce{CaFe/MnSi2O6} could display high theoretical energy densities in Ca-based batteries\cite{Torres2019}, whereas \ce{LiFeSi2O6} has been shown to undergo a reversible electrochemical reaction against \ce{Li}, thus exhibiting potential for use as an electrode material in Li-ion batteries\cite{Zhou_2014}. Recently published works have studied pyroxenes bearing Co at the \textit{M} site, with these materials having been found to be a good platform for studying the dynamically intertwined lattice, orbital, charge, and spin degrees of freedom in the quantum regime\cite{CoPyroJin}, as well as for the Kitaev model\cite{CoPyroPavel}. The magnetic properties of \ce{CaMnGe2O6} were studied by Temnikov \textit{et al.}~\cite{Temnikov2019} who computed the exchange parameters and showed that the magnetic frustration in the system is weak, which could explain the frequent occurrence of commensurate collinear antiferromagnetic structures in \ce{Ca\textsuperscript{2+}}-bearing pyroxenes.
Structural, mechanical, electronic, optical, and thermal properties of \ce{CaTGe2O6} (T = Mn, Fe, Co) were studied by Akter \textit{et al.}~\cite{CMGOAkter}
, who, based on their findings, suggested that these systems may prove useful for spintronics or optoelectronics applications in the future.
Lastly, Fakhera \textit{et al.}\cite{SMGOFakhera} studied the \ce{SrYGe2O6} (Y = Mn, Fe, Co) clinopyroxenes, studying their structural, mechanical, magnetic, and optoelectronic properties, determining based on the properties found that these point to a possible use in the context of spintronics or optoelectronics.
When the \textit{M}-site is occupied by a magnetic cation, such an arrangement leads to low-dimensional magnetic properties and to magnetic frustration, due to the competition between interchain and intrachain interactions~\cite{Jodlauk2007}. It is the existence and the possible interplay between the low dimensionality and frustration that is thought to give rise to the aforementioned diversity of magnetic properties in these materials~\cite{Ding_PRB2016}. Moreover, it has been suggested that the magnetic frustration can lead to spin spiral structures, which may be favourable towards magnetically driven ferroelectricity, making these compounds good candidates for multiferroic behaviour~\cite{Jodlauk2007}. 
The \ce{\textit{A}MnGe2O6} germanate clinopyroxene series is here explored, with the $A$-site occupied by Group-II elements from the periodic table, namely Be, Mg, Ca, and Sr. 
\textit{Ab-initio} calculations in the DFT framework performed here demonstrated that the compounds containing Be and Mg are not stable, explaining the absence of experimental results in the literature.
Further examination of compositions such as \ce{(Ca, Sr)_{1-x}Cd_xMnGe2O6} and\ce{(Ca, Sr)Mn_{1-x}Cd_xGe2O6} (where $x=0.125, 0.25$) was first conducted using \textit{ab-initio} computational simulations, which enabled the prediction of the phase stability of these alloys. Subsequently, Time-Differential (TDPAC) spectroscopy, a nuclear local probe technique, was used, after Cd ion implantation, to infer the electric field gradient (EFG) at a given Cd nuclear site. Additionally, DFT simulations were conducted to determine the local environments of the Cd probes in the \ce{CaMnGe2O6} and \ce{SrMnGe2O6} systems. After experimentally confirming that Cd can occupy both the Ca/Sr and Mn sites, the band gaps of the new alloys were estimated,  using both DFT with an on-site Hubbard $U$ and the hybrid Heyd–Scuseria–Ernzerhof (HSE06) functional;  the effective masses of the electrons and holes were also analysed. 
For the first time, we report the optical absorption and band gap of germanate clinopyroxenes through diffuse reflectance spectroscopy and compare the results with advanced DFT calculations. 
In order to investigate the fundamental effect of atomic substitution in the \ce{CaMnGe2O6} and \ce{SrMnGe2O6} systems by atoms of similar ionic radius, but distinct ionic structure, we used the insight gained from the TDPAC and DFT results to synthesize the Cd-doped germanate clinopyroxenes. This led to the first-ever study of these compounds, further expanding the understanding of their optical properties.
The diffuse reflectance measurements show the possibility of tuning the band gap through Cd doping, highlighting its potential applications in various fields.

\section{Methods}
\label{Sec:Methods}

\subsection{Computational Methodology}

\begin{figure*}[htb]
\centering
\includegraphics[width=16cm]{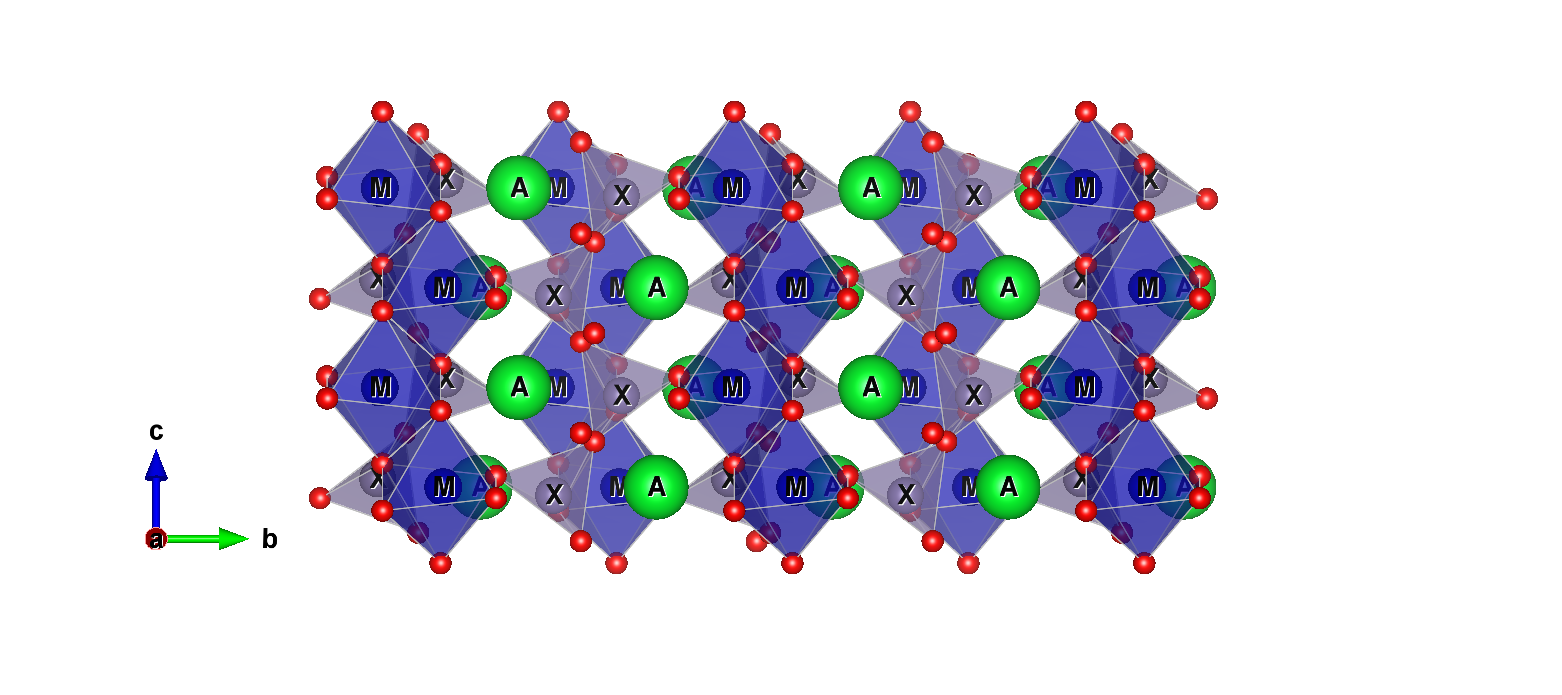}

\caption{\RaggedRight Representation of the \ce{\textit{A}MnGe2O6} clinopyroxene structure. Red spheres represent the oxygen sites, green spheres the \ce{\textit{A}} cation sites, grey tetrahedra the \ce{Ge}-site coordination polyhedra, and blue octahedra the \ce{Mn}-site coordination polyhedra.}

\label{Fig:clinopyroxene}
\end{figure*}

The ground state structure for the \ce{\textit{A}MnGe2O6} clinopyroxenes is a monoclinic Bravais lattice with space group $C2/c$ (\#15), where the Mn-cations are in octahedral sites (Wyckoff 4e atomic positions), the $A$-cations are in 8-coordinated sites (Wyckoff 4e atomic positions) and the Ge-cations are in tetrahedral sites (Wyckoff 8f atomic positions). Their structure is characterised by quasi-one-dimensional chains with edge-sharing \ce{MnO6} octahedra running along the \textit{c}-axis, and connected by corner-sharing \ce{GeO4} tetrahedra, along the \textit{c}-axis, as schematically represented in Fig.\ref{Fig:clinopyroxene}.

DFT calculations were performed by using the Quantum Espresso (QE) suite \cite{QE-2009, QE-2017},  with Projector-Augmented Wave (PAW) datasets \cite{PhysRevB.50.17953,DALCORSO2014337}. The generalized gradient approximation (GGA) with Perdew-Burke-Ernzerhof  (PBE) parametrization \cite{Perdew1996} was used to describe the exchange-correlation (XC) functional. The atomic valence configurations  were: Be[2s\textsuperscript{2}], Mg[2s\textsuperscript{2}2p\textsuperscript{6}3s\textsuperscript{2}], Ca[3s\textsuperscript{2}3p\textsuperscript{6}4s\textsuperscript{2}], Sr[4s\textsuperscript{2}4p\textsuperscript{6}5s\textsuperscript{2}], Mn[3s\textsuperscript{2}3p\textsuperscript{6}3d\textsuperscript{5}4s\textsuperscript{2}], 
Ge[3d\textsuperscript{10}4s\textsuperscript{2}4p\textsuperscript{2}],  O[2s\textsuperscript{2}2p\textsuperscript{4}], and 
Cd[4d\textsuperscript{9.5}5s\textsuperscript{2}5p\textsuperscript{0.5}]. 
The kinetic energy cutoffs were  70 and 540 Ry for, respectively, the wavefunctions and charge densities, values that were raised to 113 and 900 Ry for the EFGs calculations. The Brillouin-zone (BZ) was sampled by a $(9 \times 9 \times 18)$ Monkhorst-Pack \textbf{k}-point grid. In order to treat the strongly correlated Mn-$3d$ states, an effective on-site Hubbard $U\!=\!4$ eV potential was considered \cite{MP,doi:10.1063/1.4812323, PhysRevB.71.035105}, and this is the value used throughout this paper whenever PBE+$U$ calculations are mentioned. For an accurate description of the energy band gap values, we employed the hybrid HSE06 functional \cite{PhysRevB.80.115205} at the relaxed structures obtained from the PBE+\textit{U} calculations. 
Ground-state properties were obtained through total energy minimizations with respect to both atomic positions and cell parameters. Forces in any ion were converged to lower than 0.05 eV/Å. 
Variable cell shape relaxations were performed  using the damped Beeman ionic dynamics and the Wentzcovitch extended Lagrangian~
\cite{Wentzcovitch-PhysRevB-44-2358-1991,PhysRevB.44.5148, PhysRevLett.70.3947}.
The \ce{\textit{A}MnGe2O6} enthalpies of formation were determined by the DFT total energies of the compounds and the respective total energies of the 
\ce{\textit{A}GeO3} and \ce{MnGeO3} stable constituent oxides accordingly to Eq. \ref{eq:formation} in Appendix  \ref{Sec:A1}, and  the results shown in table \ref{table:energies} demonstrated that only  the Ca- and Sr-related compounds are stable.  
 To simulate the Cd substitution on either \textit{A} or Mn cation sites, when necessary, an $1 \times 1 \times 2$ supercell was constructed and a $(9 \times 9 \times 9)$ \textbf{k}-point grid was used to sample the BZ. 
The  EFGs, computed by the Gauge Including Projector Augmented Waves (GIPAW)\cite{CHARPENTIER20111} package routine implemented in QE, were evaluated at the Sr, Ca, Mn, Ge, O, and Cd nuclear sites.

\subsection{Experimental Methods}

\subsubsection{Ion implantation and Time-Differential Perturbed Angular Correlation Spectroscopy}

To infer if Cd can effectively substitute a regular position in the crystal structures, ion implantation followed by EFG measurements can be used. The experimental values of the EFG can be obtained via Time-Differential Perturbed Angular Correlation (TDPAC) spectroscopy. For this purpose, pelletised samples of \ce{SrMnGe2O6} and \ce{CaMnGe2O6} materials, from 
 references [\onlinecite{Ding_JMCC2016}] and  [\onlinecite{Ding_PRB2016}] were implanted at ISOLDE-CERN by a 30 keV beam of \textsuperscript{111m}Cd probes (\textsuperscript{111m}Cd$\,\rightarrow$\textsuperscript{111}Cd, $t_{1/2}$ = 48.6 minutes), which have a nuclear spin $I=5/2$ and an electric quadrupole moment  $Q$ = 0.83(13) b  in the intermediate state of decay \cite{Raghavan1989}. Small doses of the order of 10\textsuperscript{11} atoms/cm\textsuperscript{2} were implanted and subsequently an annealing was performed at a temperature of $\sim$1123~K for 20 minutes in air to recover from implantation damage. Confirmation for the recovery of point defects and for the incorporation of \ce{Cd} into the lattice was given by the TDPAC measurements themselves, which yielded well-defined experimental \textit{R(t)} anisotropy functions. The measurements were performed on 6-\ce{BaF2} detectors TDPAC spectrometers \cite{Butz1989}, equipped with either a closed-cycle cryostat or a special high-temperature furnace for temperature control. 
The experimental \textit{R(t)} function was fitted with exact numerical methods that build the expected observable by solving the hyperfine interaction Hamiltonian's characteristic equations \cite{Schatz1996,PhysRevB.73.100408,PhysRevB.47.8763,NNFITBarr,NNFIT}. 
The EFG, that characterizes the surrounding charge density that interacts with a
probe nucleus, can be represented as a rank-2 symmetric and traceless tensor. 
As such it is always possible to be diagonalized into the principal axis system. The experimentally observable EFG is conventionally described by only two parameters namely: the principal component $V_{zz}$ and the 
asymmetry parameter $\eta$ defined as
\begin{equation}
    \eta=\frac{V_{xx}-V_{yy}}{V_{zz}}.
\end{equation} 
\noindent
The EFG principal axis tensor components $V_{zz}$, $V_{yy}$, and $V_{xx}$ are defined such that $|V_{zz}|\geq|V_{yy}|\geq|V_{xx}|$ \cite{Schatz1996}.

The signature for the interaction between the probe atoms and their local environment is the perturbation function $G_{kk}(t)$. In the case of static electric quadrupole interactions, the interaction can be written as \cite{doi:10.1080/00150199308008701}:
\begin{equation}
 \begin{aligned}
G_{kk}(t)=\sum_{n} s_{k_{n}}\cos(\omega_{n}t)e^{-\delta\omega_{n}t/2},
 \end{aligned}
 \end{equation}
where $s_{k_{n}}$ is a function of $\eta$ and $\omega_{n}$ is a function of  both $\eta$ and the fundamental frequency $\omega_{Q}$. The exponential term arises from randomly distributed defects and lattice strains that result in an attenuation of the experimental \textit{R(t)} function. In the case of the \textsuperscript{111m}Cd probes, the intermediate level in the decay has a nuclear spin of $I=5/2$, which leads to a split of the intermediate level into three sub-levels due to the quadrupole interaction and three transition frequencies, $\omega_{1}$, $\omega_{2}$, and $\omega_{3}=\omega_{1}+\omega_{2}$. The fundamental frequency is usually defined as \cite{Schatz1996}:
\begin{equation}
 \begin{aligned}
\omega_{Q}=\frac{eQV_{zz}}{4I(2I-1)\hbar},
 \end{aligned}
 \end{equation}
 where $Q$ is the nuclear quadrupole moment, $e$ is the electron charge, and $\hbar$ is the reduced Planck constant.
 In the case of static interactions, the experimental anisotropy function can be written as:
 \begin{equation}
     \begin{aligned}
         R(t)=\sum{A_{kk}G_{kk}(t)}
     \end{aligned}
 \end{equation}
where $A_{kk}$ are the anisotropy coefficients for the nuclear decay cascade.

\subsubsection{Sample Synthesis}
Proceeding akin to what had previously been reported for the parent compounds, \ce{CaMnGe2O6} \cite{Ding_PRB2016} and \ce{SrMnGe2O6} \cite{Ding_JMCC2016}, and guided by DFT calculations and TDPAC measurements, 
which indicate the stability of these compounds upon Cd doping at the Ca/Sr and Mn-sites, the \ce{Ca_{1-x}Cd_{x}MnGe2O6}, \ce{CaMn_{1-x}Cd_{x}Ge2O6}, \ce{Sr_{1-x}Cd_{x}MnGe2O6},  and \ce{SrMn_{1-x}Cd_{x}Ge2O6}  compounds ($x=0.125, 0.25$) were synthesised following the solid-state reaction method. For the Ca/Sr-based clinopyroxenes, stoichiometric quantities of \ce{CaCO3} (99 \%)/\ce{SrCO3} (99.995 \%), \ce{CdO} (99.99 \%), \ce{MnO} (99 \%), and \ce{GeO2} (99.998 \%) were hand-ground together and the powder fired in air at 1373 K for 50 hours with a heating and cooling rate of 10 K/min. The resulting powder was re-ground and pelletised, then sintered for 12 hours under the same conditions for Sr-based clinopyroxenes and at 1073 K for 4 hours in air for Ca-based clinopyroxenes. Pristine samples were synthesised following references [\onlinecite{Ding_PRB2016}] and [\onlinecite{Ding_JMCC2016}].
\par
The phase purity of the synthesised samples were assessed by monitoring the powder X-ray diffraction (PXRD) data collected using Rigaku SmartLab diffractometer (45 kV, 200 mA) operating with Cu K$\alpha$ in Bragg-Brentano geometry. Crystal structure analysis was performed using the Fullprof software \cite{Fullprof1993,Fullprof2001,Rodriguez1990}, the corresponding structural refinement plots and structural parameters are provided in Appendix \ref{App:Xray}.

\subsubsection{Reflectometry}

In order to inspect the band gap widths, diffuse reflectance measurements were acquired using CARY 50 Varian spectrophotometer in a range from 200 to 1000 nm, and the \ce{BaSO4} standard compound as the white background reference. The acquired spectrum was converted using the Kubelka-Munk function \cite{Kubelka}, where the magnitude $F(R_{\infty})$ is proportional to the absorption coefficient ($\alpha$). The optical band gap ($E_{g}$) is calculated following the relation presented by Tauc and expressed by Davis \& Mott: $(\alpha\cdot E)^{\frac{1}{n}} = (h\cdot \nu – E_{g})$, where $E = h\cdot \nu$ is the photon energy and $E_{g}$ is the optical band gap energy. The power-law exponent, $n$, depends on the transition type: $n = \frac{1}{2}$ for a direct $E_{g}$ and $n = 2$ for an indirect $E_{g}$. The value of indirect or direct $E_{g}$ is estimated by plotting $(\alpha\cdot h \cdot \nu)^{\frac{1}{2}}$ or $(\alpha\cdot h\cdot\nu)^{2}$, respectively, as a function of the photon energy and extrapolating it to $\alpha=0$ \cite{ReflectometryMakula}.

\section{Results and Discussion}

\subsection{Density Functional Theory Calculations}

To ascertain the possibility of synthesing other compounds in the \ce{\textit{A}MnGe2O6} series ($A\!=\,$Be, Mg, Ca, Sr), the enthalpy of formation from constituent oxides of the series has been examined through DFT calculations. As mentioned before, we have determined that only the \ce{CaMnGe2O6} and \ce{SrMnGe2O6} clinopyroxene compounds are thermodynamically stable. This is evidenced by their, respective, -0.19  and -0.11 eV/f.u. (negative) enthalpies of formation, calculated through equation \ref{eq:formation}  (Appendix \ref{Sec:A1}). On the other hand, both  \ce{BeMnGe2O6} and \ce{MgMnGe2O6} presented  0.12 eV/f.u. (positive) enthalpies of formation, supporting that these are not stable against dissociation into the respective stable constituent oxides: \ce{BeGeO3} or \ce{MgGeO3} and \ce{MnGeO3}. These results are consistent with the fact that only the former two compounds have been experimentally synthesised, whereas the synthesis of the latter two has not yet been reported. Such observation implies that different reaction pathways towards synthesis will have still to be explored in order to possibly realize experimentally the compounds containing Be and Mg.

\begin{table*}[htb]
\caption{\RaggedRight Experimental~\cite{Ding_PRB2016, Ding_JMCC2016} and theoretical lattice parameters for SrMnGe$_2$O$_6$ and CaMnGe$_2$O$_6$  compounds obtained using the PBE and PBE+$U$ approximations.
}
 \label{table:latparam}
\centering

		\begin{tabular}{llcccc}
			\hline \hline
&\multicolumn{1}{l|}{}&&&& \\[-3mm]
\multicolumn{1}{l}{~~System} & \multicolumn{1}{l|}{Research type~~} &  $a\,$(\AA) & $b\,$(\AA) & $c\,$(\AA) & $\beta\,(^{\circ})$  \\
&\multicolumn{1}{l|}{}&&&& \\[-3mm]
			\hline
 &\multicolumn{1}{l|}{}&&&& \\[-3mm]  
 \multicolumn{1}{l}{\multirow{3}{*}{~~SrMnGe$_2$O$_6$~~~}}   &    \multicolumn{1}{l|}{Expt.} & ~~10.3511(6)~~ & ~~9.4204(5)~~ & ~~5.5093(3)~~ & ~~104.700(2)~~ \\
 &\multicolumn{1}{l|}{}&&&& \\[-3.5mm]  
  \multicolumn{1}{l}{}        & \multicolumn{1}{l|}{Theo. PBE}  & 10.5295 & 9.4912 & 5.5992 & 105.029 \\
   &\multicolumn{1}{l|}{}&&&& \\[-3.5mm]  
    \multicolumn{1}{l}{}        & \multicolumn{1}{l|}{Theo. PBE+$U$~~~} & 10.5638 & 9.5795 & 5.6230 & 104.906 \\ 
    \hline
     &\multicolumn{1}{l|}{}&&&& \\[-3mm] 
     \multicolumn{1}{l}{\multirow{3}{*}{~~CaMnGe$_2$O$_6$ }}   &    \multicolumn{1}{l|}{Expt.} & 10.2794(3) & 9.1756(3) & 5.4714(2) & 104.244(2) \\
      &\multicolumn{1}{l|}{}&&&& \\[-3.5mm]  
  \multicolumn{1}{l}{}        & \multicolumn{1}{l|}{Theo. PBE}  & 10.4006 & 9.2248 & 5.5194 & 103.921 \\
   &\multicolumn{1}{l|}{}&&&& \\[-3.5mm]  
    \multicolumn{1}{l}{}        & \multicolumn{1}{l|}{Theo. PBE$+U$} & 10.4495 & 9.3204 &  5.5472 & 104.109\\
\hline \hline
\end{tabular}

\end{table*}

Table \ref{table:latparam} summarizes the calculated lattice parameters for the \ce{SrMnGe2O6} and \ce{CaMnGe2O6} systems, computed with both the PBE and PBE$+U$ (using $U\!=\!4$ eV) approximations, as well as the experimental values  reported in the literature\cite{Ding_PRB2016, Ding_JMCC2016}. 
Regarding the \ce{SrMnGe2O6} system, we can observe that 
the lattice parameters computed with the PBE functional are overestimated, with respect to the experimental values, corresponding to a relative increase of $\sim$1.7\%, $\sim$0.8\%, and $\sim$1.6\%, respectively,  for \textit{a}, \textit{b}, and \textit{c} parameters, which is in line with the well-known tendency of the PBE functional \cite{PhysRevB.79.085104}. As compared to the lattice parameters likewise computed with the PBE functional for this system by Fakhera \textit{et al.} \cite{SMGOFakhera}, the values we obtained for the lattice parameters are 0.6-1.2\% smaller, with the difference likely being owed to the difference in magnetic ordering, which in our case was considered as antiferromagnetic (AFM), as is observed experimentally, whereas the aforementioned work considered a 
ferromagnetic (FM) configuration for their calculations. The inclusion of the on-site effective Hubbard-\textit{U} potential results in a further relative increase of these parameters with $\sim$2.1\%, $\sim$1.7\%, and $\sim$2.1\%, when compared to the experimental values. Though the $\beta$ monoclinic angle is overestimated for both PBE and PBE$+U$ approximations, when comparing to experimental data, the difference is in fact smaller when considering the PBE$+U$ calculations, resulting in a value closer to the experimental one.
In the case of \ce{CaMnGe2O6}, the \textit{a} and \textit{b} theoretical lattice parameters obtained from PBE calculations likewise are overestimated, relative to experimental data, by $\sim$1.2\% and $\sim$0.5\%, respectively. On the other hand we observed that the \textit{c} parameter is underestimated  by $\sim$0.9\%. The lattice parameters calculated with the PBE functional by Akter \textit{et al.} \cite{CMGOAkter} are in quite good agreement with the ones we found, differing by less than 0.1\%, with the exception of the \textit{b} parameters, which is 1.1\% larger than ours.  Analogously to what was observed for \ce{SrMnGe2O6}, the $a$ and $b$ lattice parameters that resulted from the PBE$+U$ approach display an increase with respect to the PBE values, presenting an overestimation of $\sim$1.7\% and $\sim$1.6\%, respectively,  relative to the experimental results; whereas \textit{c} parameter is still underestimated, though smaller than the one obtained by the PBE functional, of  $\sim$0.4\%. Also, and distinctly from what was observed for \ce{SrMnGe2O6}, the theoretical monoclinic angle is underestimated in relation to the experimental angle, though again, results from the PBE$+U$ approximation are closer to the experimental value.

\begin{figure*}[htb]
 \centering
    \includegraphics[width=0.9\linewidth]{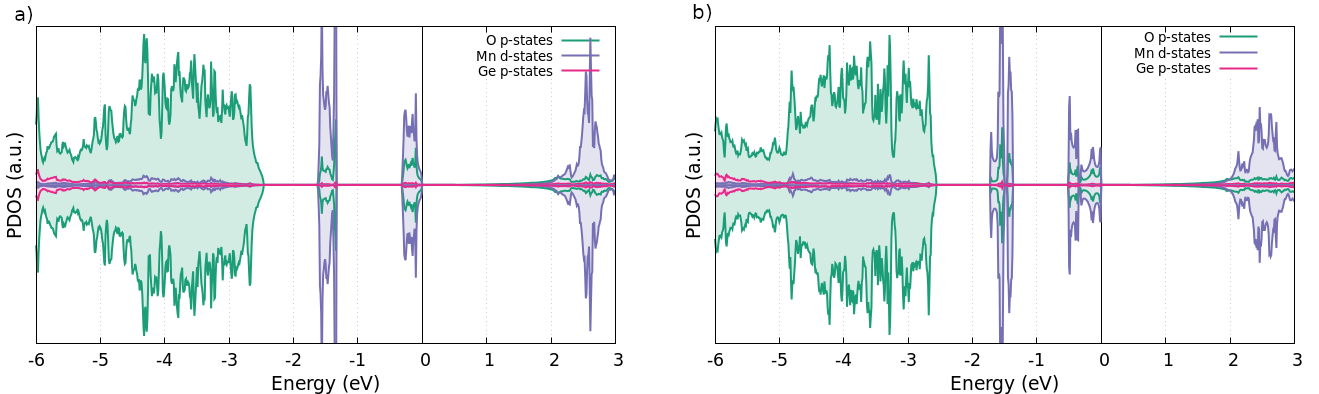}
\caption{\RaggedRight Projected density of states for (a) \ce{CaMnGe2O6}  and (b) \ce{SrMnGe2O6}, computed with 
the PBE approximation. The Fermi level is set to zero and corresponds to the VBM.}
\label{fig:DOSU0}
\end{figure*}

The \ce{CaMnGe2O6} and \ce{SrMnGe2O6}  projected density of states (PDOS), obtained considering the PBE calculations (without the Hubbard-$U$ correction) 
were computed and are displayed in figure \ref{fig:DOSU0}, in which the valence band maximum (VBM) are aligned at the Fermi energy and are set to zero.
Two narrow bands form the  VBM region,  being mainly composed by the split of the  Mn-$3d$ derived bonding states, hybridized with a smaller contribution from the O-$p$ states. Energetically below these narrow bands ($<$ -2.5 eV), a broader band is observed and it is mostly dominated by the O-$p$ states, with a very small mixture of Mn-$3d$ states. The conduction band minimum (CBM) region, on the other hand, presents low density of states and hybridization between the O-$p$ and Ge-$p$ states. For increasing energies, the DOS significantly increases forming a narrow resonant band, between 2.0 and 3.0 eV, that is mainly composed by the unoccupied Mn-$3d$ states. For both studied systems the PDOS are qualitatively very similar, although at the conduction band region the observed Mn resonant peaks for \ce{SrMnGe2O6} are broader when compared to those of \ce{CaMnGe2O6}.
The electronic band gaps are also noted to be fairly narrow, being 0.46 eV and 0.51 eV, respectively, for \ce{CaMnGe2O6} and \ce{SrMnGe2O6}, which are narrower than what have been reported for similar Si-based compounds \cite{PhysRevB.81.045118,https://doi.org/10.1002/cphc.201701155}.

\begin{figure*}[htb]
\centering
    \includegraphics[width=0.93\linewidth]{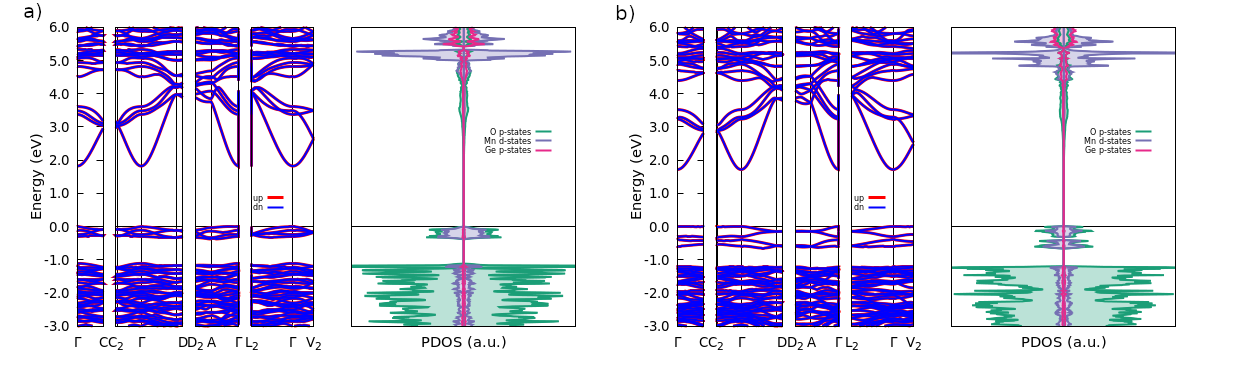}
    \caption{\RaggedRight a) \ce{CaMnGe2O6}  and b) \ce{SrMnGe2O6} band structures  in the high-symmetry directions of the BZ (left)
    and projected density of states (right), computed with 
    PBE$+U$ and antiferromagnetic ordering. 
    The Fermi level is set 
to zero and matches the VBM.}
\label{fig:DOS}
\end{figure*}

\begin{figure}[tbp]
    \centering
\includegraphics[width=0.55\linewidth]{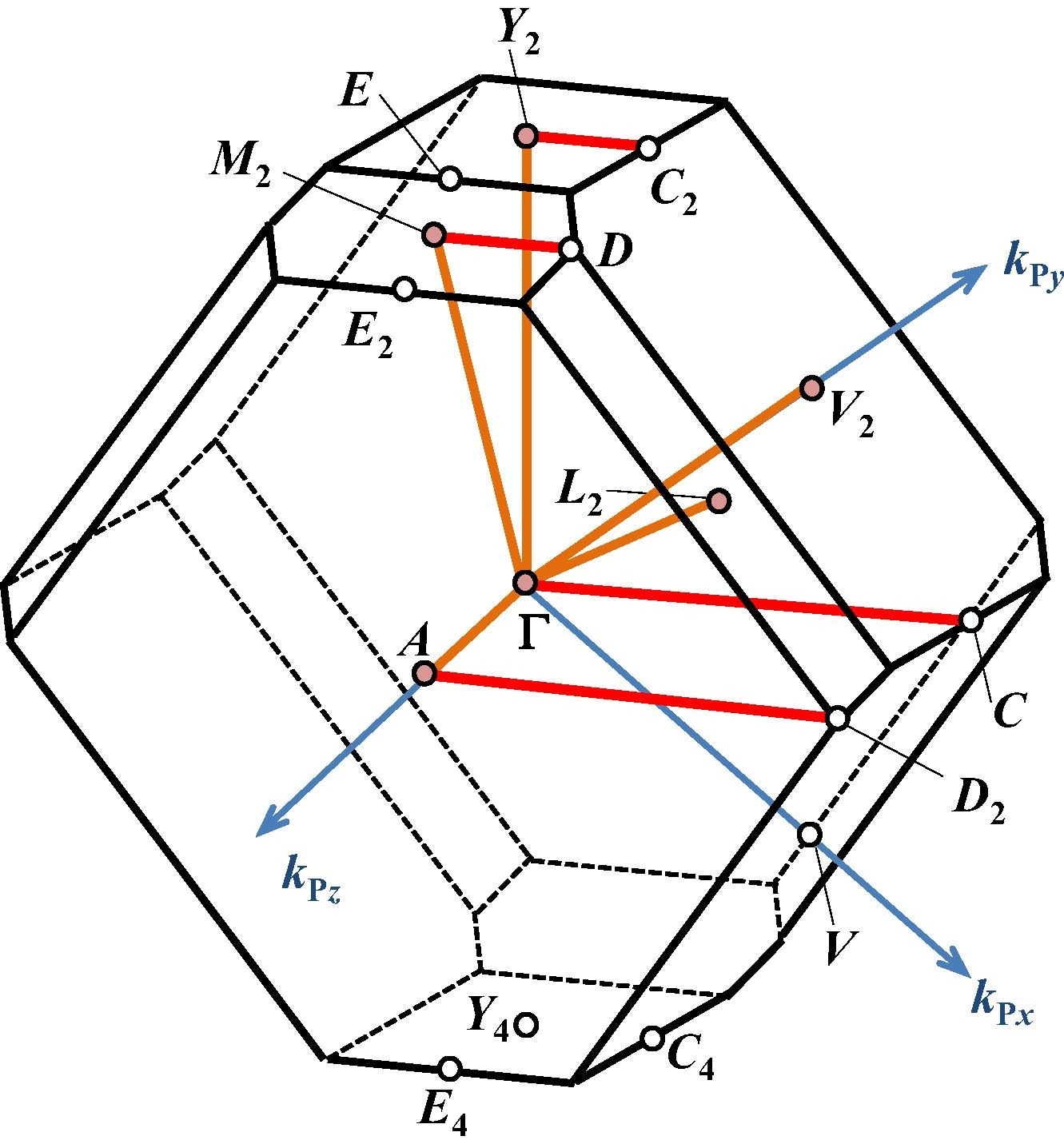}
\caption{\RaggedRight Base-centered monoclinic lattice first Brillouin zone  and respective high-symmetry points and directions. Adapted from reference [\onlinecite{HINUMA2017140}].} 
\label{fig:lattice}
\end{figure}

Since the band gap widths of these systems are likely underestimated in the absence of the Hubbard-$U$ correction, we have also considered the PDOS by including the $U\!=\!4$ eV correction on the Mn-$3d$ states. These are shown in figure \ref{fig:DOS} along with the respective electronic band structures in a path along the high-symmetry points and directions of the  BZ depicted in figure \ref{fig:lattice}, generated with the SeeK-path tool. \cite{HINUMA2017140,togo2024textttspglibsoftwarelibrarycrystal}


When including the $U$ parameter and comparing to the PDOS results without this correction,  differences are mainly evidenced at the VBM region, where the Mn-$3d$ derived two narrow sets of bands observed in figure \ref{fig:DOSU0}, now overlap and form one narrow set of bands for \ce{CaMnGe2O6} (figure \ref{fig:DOS} a)). As for the \ce{SrMnGe2O6} system, the narrow bands are now energetically very close, almost forming one narrow band (figure \ref{fig:DOS} b)). Moreover, these narrow bands are now dominated by a stronger hybridization between Mn-$3d$ and O-$p$ states. Since the inclusion of the $U$ parameter localizes and positions the unoccupied narrow Mn-$3d$ states higher in energy (between 5.0 and 5.5 eV), the bands around the CBM show slightly increased $p$ hybridized states of O and Ge than when $U$ correction is not considered. 

The PBE$+U$ theoretical band gaps are wider when compared to the previous case, with 1.82 and 1.70 eV for the \ce{CaMnGe2O6} and \ce{SrMnGe2O6} compounds, respectively, as can be observed in table \ref{table:gaps}. Despite this, the gap widths are still below the range of other similar Si-based clinopyroxenes \cite{PhysRevB.81.045118,https://doi.org/10.1002/cphc.201701155}. We must stress that subtleties persist with respect to the analysis of the VBM, since the valence bands are quite flat, and thus the electronic bands along the different high symmetry segments/points are energetically very close.  
Comparing to the band gaps reported in the theoretical works through GGA+U by Fakhera \textit{et al.} \cite{SMGOFakhera} and Akter \textit{et al.} \cite{CMGOAkter}, they both report a different nature for the band gaps, with both finding their respective studied systems as half-metallic and with significantly different widths for the band-gaps (Akter \textit{et al.} report a 3.05 eV half-metallic gap for \ce{CaMnGe2O6}, whilst Fakhera \textit{et al.} report a direct half-metallic gap for \ce{SrMnGe2O6}), whereas we found the systems to be insulating. This is likely a result of their choice to perform the calculations with an FM configuration for the systems, whereas the systems are known experimentally to have an AFM ground state, moreover, the AFM configuration is also found to be more stable in the context of DFT calculations. This conclusion is further supported by the work
reported by Temnikov \textit{et al.} \cite{Temnikov2019}, who performed DFT simulations on the \ce{CaMnGe2O6} system considering an AFM alignment and, as with the present work, reported an insulating character for the system with a bandgap of $\sim$2 eV, similar to that obtained here.

From experimental in-house diffuse reflectance measurements we have obtained the optical indirect energy band gap widths of 3.43 and 3.38 eV  for \ce{CaMnGe2O6} and \ce{SrMnGe2O6}, respectively. Therefore, we have also computed the band gap widths by employing the hybrid HSE06 approach for the exchange-correlation potential. These values are 2.94 and 2.91 eV, for \ce{CaMnGe2O6} and \ce{SrMnGe2O6}, respectively, in much better agreement with experimental data (table \ref{table:gaps}).

From figure \ref{fig:DOS}, one can observe that the VBM 
is much flatter than the CBM, 
which shows narrow parabolas around the zone-centre, indicating that the holes effective mass, $m^{*}_{h}$,  is much larger than the electrons effective mass, $m^{*}_{e}$. Such an observation is an indication of a higher mobility of the electrons comparatively to the holes of the systems. Since the charge carriers effective mass  is an essential parameter for the performance of photoelectric and photo-catalytic activities \cite{LIU20161}, and also for semiconductor radiation detectors \cite{Lee_JElectronMater_28_766}, we have considered further analysing respective properties (table \ref{table:gaps}). Therefore and to quantify the effective masses of holes and electrons, we employ the well-known definition of the effective mass as
\begin{equation}
    m^{*}=\frac{\hbar^{2}}{\frac{\partial^2 E}{\partial k^2}},
\end{equation}

\noindent\noindent with the second derivative of the electronic dispersion relation being obtained from a quadratic fit around the energy range where the band edges are located.

In the case of \ce{CaMnGe2O6}, we obtain a $m^{*}_{e}$,  around the $\Gamma$-point, very close to that of the free electrons, with $m^{*}_{e}=1.028 m_{0}$ (where $m_{0}$ is the free electron mass of  9.11$\times$10$^{-31}$ kg). While for the hole carriers, and as expected by considering the flatness-shape of the VBM, the mass of holes proved to be substantially larger (in absolute terms) with $m^{*}_{h}=26.125 m_{0}$, calculated along the $\overline{A\Gamma} \rightarrow A$  high-symmetry segment (table \ref{table:gaps}).

\begin{table*}[t!]
\caption{\RaggedRight Theoretical electronic indirect band gaps  ($E_g$) and charge carriers effective masses $m^{*}_{e}$ and $m^{*}_{h}$  (in units of the electron mass $m_{0}=9.11\times 10^{-31}$ kg)  for the pristine and Cd-doped systems obtained with PBE+$U$. Values shown in shaded cells are those obtained with the HSE06 hybrid exchange-correlation functional. Experimental optical direct (dir.) and indirect (ind.) band gaps are also shown.}

\label{table:gaps}

\begin{center}
 \begin{tabular}{|l|c|c|c|c|c|}
\hline\hline
~~~~~~~~System  & Theo. $E_g$ (eV) & Expt. ind. $E_g$ (eV) & Expt. dir. $E_g$ (eV) &~~$m^{*}_{e}$ ($m_{0}$)~~ & $m^{*}_{h}$ ($m_{0}$)\\
\hline
&&&& &\\ [-2mm]
\multirow{2}*{\ce{SrMnGe2O6}}  & 1.70 & \multirow{2}*{3.39(1)} & \multirow{2}*{3.66(2)} & \multirow{2}*{1.023} & \multirow{2}*{9.380 ($\Gamma \rightarrow D$)} \\
  & \cellcolor{lightgray}2.91 & & & & \\ 
&&&& &\\ [-2mm]
\hline
&&&& &\\ [-2mm]
\ce{Sr_{0.875}Cd_{0.125}MnGe2O6}  & 1.55  & 3.28(6) & 3.54(2) & 1.028 & 58.223 ($\overline{C_{2}\Gamma} \rightarrow C_{2}$)\\
&&&& &\\ [-2mm]
\hline
&&&& &\\ [-2mm]

 \multirow{2}*{\ce{Sr_{0.75}Cd_{0.25}MnGe2O6}}  & 1.49  & \multirow{2}*{3.10(5)} & \multirow{2}*{3.33(2)} & \multirow{2}*{1.106} & \multirow{2}*{58.287 ($\overline{C_{2}\Gamma} \rightarrow C_{2}$)}\\
   & \cellcolor{lightgray}2.60 & & & & \\
&&&& &\\ [-2mm]
\hline
&&&& &\\ [-2mm]
\ce{SrMn_{0.875}Cd_{0.125}Ge2O6}  & 1.56  & 3.33(2) & 3.54(2) & 1.057 & 40.062 ($\overline{C_{2}\Gamma} \rightarrow C_{2}$)\\
&&&& &\\ [-2mm]
\hline
&&&& &\\ [-2mm]
\multirow{2}*{\ce{SrMn_{0.75}Cd_{0.25}Ge2O6}}  & 1.50 & \multirow{2}*{3.18(3)} & \multirow{2}*{3.59(3)} &\multirow{2}*{1.068} & \multirow{2}*{46.882 ($V_{2} \rightarrow \Gamma)$}\\
& \cellcolor{lightgray}2.66 &  & & & \\ 
&&&& &\\ [-2mm]
\hline \hline
&&&& &\\ [-2mm]
\multirow{2}*{\ce{CaMnGe2O6}}   & 1.82 & \multirow{2}*{3.44(1)} & \multirow{2}*{3.78(2)} & \multirow{2}*{1.028} & \multirow{2}*{26.125 ($\overline{A\Gamma} \rightarrow A$)}\\
  & \cellcolor{lightgray}2.94 &  & & & \\ 
  &&&& &\\ [-2mm]
\hline
&&&& &\\ [-2mm]
\ce{Ca_{0.875}Cd_{0.125}MnGe2O6}  & 1.59 & 3.30(1) & 3.37(2) &1.028 & 67.562 ($\overline{A\Gamma} \rightarrow A$)\\
&&&& &\\ [-2mm]
\hline
&&&& &\\ [-2mm]
\multirow{2}*{\ce{Ca_{0.75}Cd_{0.25}MnGe2O6}}  & 1.42 & \multirow{2}*{3.26(1)} & \multirow{2}*{3.48(1)} & \multirow{2}*{0.996} & 27.875 ($\overline{A\Gamma} \rightarrow A$) \\ & \cellcolor{lightgray}2.60 & & & & 45.259 ($\overline{A\Gamma} \rightarrow \Gamma$)\\
&&&& &\\ [-2mm]
\hline
&&&& &\\ [-2mm]
\multirow{2}*{\ce{CaMn_{0.875}Cd_{0.125}Ge2O6}}  & \multirow{2}*{1.58} & \multirow{2}*{3.26(1)} & \multirow{2}*{3.37(1)} & \multirow{2}*{1.033} & 18.704 ($\overline{A\Gamma} \rightarrow A$)\\ & & & & & 68.128  ($\overline{A\Gamma} \rightarrow \Gamma$) \\
&&&& &\\ [-2mm]
\hline
&&&& &\\ [-2mm]
\multirow{2}*{\ce{CaMn_{0.75}Cd_{0.25}Ge2O6}} & 1.61 & \multirow{2}*{3.16(4)} & \multirow{2}*{3.49(3)} & \multirow{2}*{1.05} & \multirow{2}*{118.325 ($V_{2} \rightarrow \Gamma$)}\\
& \cellcolor{lightgray}2.77 & & & & \\ 
\hline\hline
\end{tabular}
\end{center}
\end{table*}

The simulation of the \ce{(Ca,Sr)MnGe2O6} supercells containing Cd substitution at either Ca/Sr or Mn sites  demonstrated that they are stable.  
The electronic band structures and the PDOS spectra were obtained for all doped systems and are displayed in figures \ref{fig:FigSrBands_DOS} to \ref{fig:Ca2Mn2Ge4O12AFM_bands} in Appendix \ref{Sec:A3}.  The band gap transitions are summarized in Table \ref{table:gaps} where it can be observed that the doped structures show narrower indirect band gaps than the respective pristine compounds.

Regarding the \ce{CaMnGe2O6} supercells containing Cd substitution at the Ca site,  we found that the electrons effective masses only yielded mild variation when compared to the pristine system. For the electrons, the $m^{*}_{e}$ values showed very little to no increase, being  $m^{*}_{e}=0.996 m_{0}$ and $m^{*}_{e}=1.028 m_{0}$, for the \ce{Ca_{0.75}Cd_{0.25}MnGe2O6} and \ce{Ca_{0.875}Cd_{0.125}MnGe2O6} systems, respectively. In the case of the holes, we observe that for the \ce{Ca_{0.75}Cd_{0.25}MnGe2O6} cell, and by comparing to the pristine system, $m^{*}_{h}$ slightly increases along the $\overline{A\Gamma} \rightarrow \Gamma$ with $m^{*}_{h}=45.259 m_{0}$, although showing similar values along $\overline{A\Gamma} \rightarrow A$ with $m^{*}_{h}=27.875 m_{0}$.

Regarding the pristine \ce{SrMnGe2O6} structure, we found a similar value for the electron effective mass with $m^{*}_{e}=1.023 m_{0}$. However, a significantly more parabolic dispersion relation at the VBM when compared to \ce{CaMnGe2O6}, leading to a hole effective mass of $m^{*}_{h}=9.380 m_{0}$, when considering the $\Gamma\rightarrow D$ segment. We must however state that since the valence band dispersion is quite asymmetric, the interpolation with a quadratic function did not yield a very accurate fit to the dispersion relation; therefore the value of the $m^{*}_{h}$ should be understood more as an order of magnitude rather than a quantitative value. 

It is observed, once again,  
that with the Cd substitution, there is only a mild variation of $m^{*}_{e}$, leading to a slightly increased value of $m^{*}_{e}=1.028 m_{0}$ for \ce{Sr_{0.875}Cd_{0.125}MnGe2O6}, and a more pronounced increase for the 
\ce{Sr_{0.75}Cd_{0.25}MnGe2O6} compound with $m^{*}_{e}=1.106 m_{0}$. As for the \ce{SrMn_{0.75}Cd_{0.25}Ge2O6} and \ce{SrMn_{0.875}Cd_{0.125}Ge2O6} systems, a slight increase of the effective masses is observed to $m^{*}_{e}=1.068 m_{0}$ and $m^{*}_{e}=1.057 m_{0}$, respectively. 
With respect to the $m^{*}_{h}$, we again note that due to the flat-shape of the bands, a parabolic interpolation does not provide a dependable quantitative value, so we can only state with confidence that this value should remain of the same order as that of the pristine \ce{SrMnGe2O6} compound. 

Concluding this analysis, we must state that such observed differences of the masses between the charge carriers is quite interesting because of the increased possibility to enhance the separation of the electron–hole pairs, and even of evidencing large electron and
hole drift lengths \cite{Lee_JElectronMater_28_766}, thus affecting the opto-electronic properties of the studied clinopyroxenes 
 \cite{C9CY00997C}.

The DFT EFG calculations, at all crystallographic positions  
were performed for the pristine \ce{{\textit{A}}MnGe2O6} systems and for the 
supercells with Cd substitution, either at the $A$-site or  Mn-site. These results, together with the ones obtained experimentally, are summarised in Table \ref{table:EFG_DFT} and discussed in the next section. A more comprehensive list of computed values can be found in Appendix \ref{Sec:A2}, Table \ref{table:EFG}. There, the EFG at each \textit{A}, Mn and Ge atomic position is presented for the pristine compounds alongside with the result for the Cd doped systems.

\begin{table}
\caption{\RaggedRight Experimental (gray shaded cells) 
and theoretical  EFG parameters for \ce{SrMnGe2O6} and \ce{CaMnGe2O6} related compounds. For the Cd compounds, theoretical values are at the Cd site.}
\label{table:EFG_DFT}
\centering

 \begin{tabular}{l|c|c}
\hline \hline
&& \\ [-3mm]
~~~~~~~~~System & {$|V_{zz}|$} (V/\AA$^{2}$)~~ & $\eta$ \\
& &  \\[-3mm]
\hline
\cellcolor{gray!30}  & 
\cellcolor{gray!30} &  
\cellcolor{gray!30}  \\
&& \\[-6.5mm]
\cellcolor{gray!30} EFG\textsuperscript{\ce{SMGO1}} (13 K) & 
\cellcolor{gray!30}120(1) & \cellcolor{gray!30}0.28(1) \\
\cellcolor{gray!30}  & 
\cellcolor{gray!30} &  
\cellcolor{gray!30}  \\
&& \\[-6.5mm]
\cellcolor{gray!30} EFG\textsuperscript{\ce{SMGO2}} (13 K) & 
\cellcolor{gray!30}20(1) & ~~~\cellcolor{gray!30}0.81(6)~~~ \\
&& \\[-2mm]
\ce{SrMnGe2O6} ~~Sr-site~~~~ & 64 & 0.54 \\
&& \\[-3mm]
\ce{Sr_{0.875}Cd_{0.125}MnGe2O6} & 108 & 0.14 \\
&& \\[-3mm]
\ce{Sr_{0.75}Cd_{0.25}MnGe2O6}~~~~ & 103 & 0.14 \\
&& \\[-1mm]
\ce{SrMnGe2O6} Mn-site & 14 & 0.32 \\
&& \\[-3mm]
\ce{SrMn_{0.875}Cd_{0.125}Ge2O6} & 21 & 0.74 \\
&& \\[-3mm]
\ce{SrMn_{0.75}Cd_{0.25}Ge2O6} & 22 & 0.67 \\
&& \\ [-2mm]
\hline
\cellcolor{gray!30}  & 
\cellcolor{gray!30} &  
\cellcolor{gray!30}  \\ 
&& \\[-6.5mm]
\cellcolor{gray!30} EFG\textsuperscript{\ce{CMGO1}} (11 K) & 
\cellcolor{gray!30}130(1) & \cellcolor{gray!30}0.15(1) \\
\cellcolor{gray!30}  & 
\cellcolor{gray!30} &  
\cellcolor{gray!30}  \\
&& \\[-6.5mm]
\cellcolor{gray!30} EFG\textsuperscript{\ce{CMGO2}} (351 K) & 
\cellcolor{gray!30}10(2) & \cellcolor{gray!30}--- \\
&& \\[-2mm]
\ce{CaMnGe2O6} ~Ca-site & 34 & 0.52 \\
&& \\[-3mm]
\ce{Ca_{0.875}Cd_{0.125}MnGe2O6} & 109 & 0.01 \\
&& \\[-3mm]
\ce{Ca_{075}Cd_{0.25}MnGe2O6} & 108 & 0.02 \\
&& \\[-1mm]
\ce{CaMnGe2O6} Mn-site & 8 & 0.49 \\
&& \\[-3mm]
\ce{CaMn_{0.825}Cd_{0.125}Ge2O6} & 12 & 0.21 \\ 
&& \\[-3mm]
\ce{CaMn_{0.75}Cd_{0.25}Ge2O6} & 12 & 0.31 
\\[2.5mm]
\hline \hline
		\end{tabular}	

\end{table}

\subsection{Time-Differential Perturbed Angular Correlation}
\begin{figure}[htb]

\centering
\includegraphics[width=1 \linewidth]{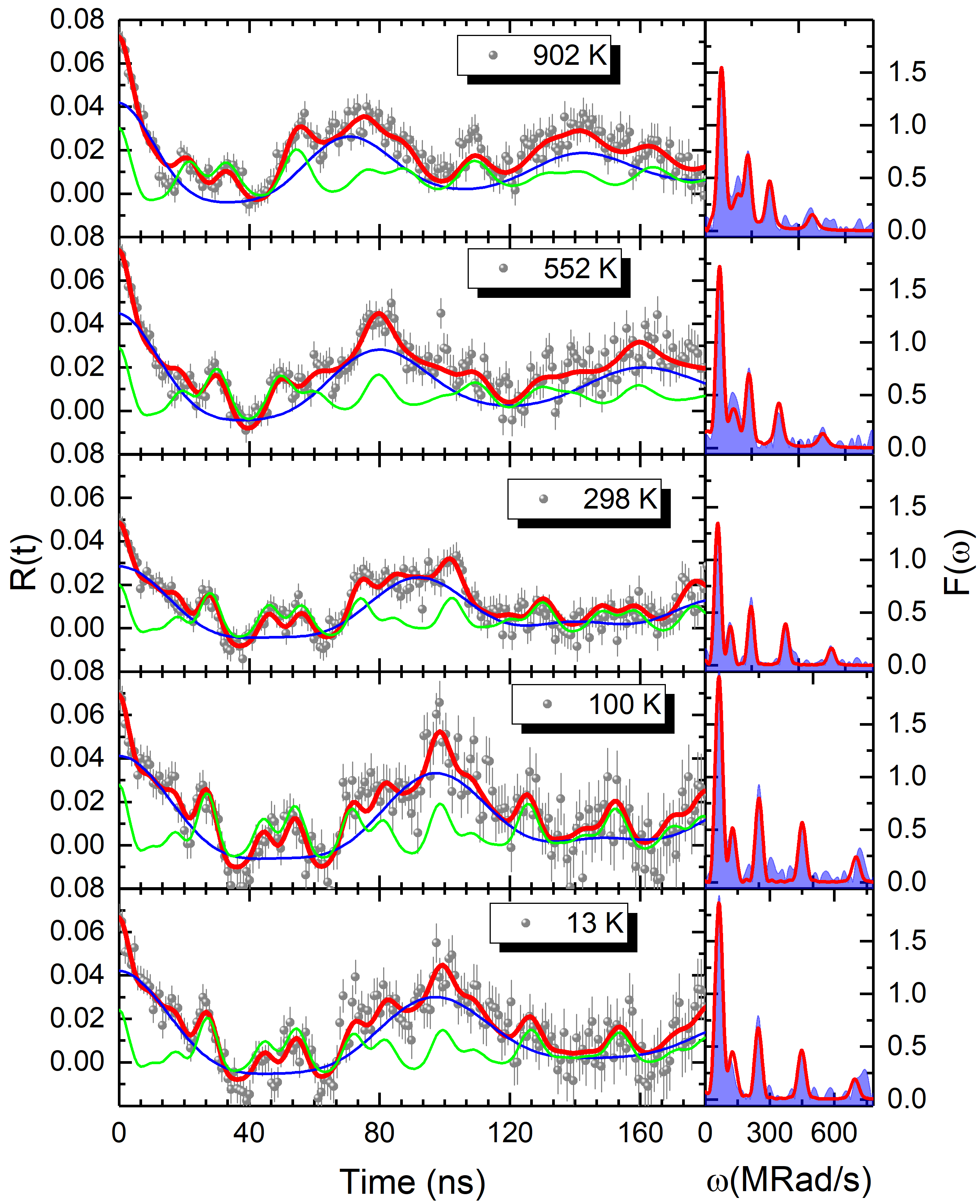}
\caption[\RaggedRight Representative \textit{R(t)} spectra of \ce{SrMnGe2O6} and corresponding fits (left), as well as Fourier transforms (right). Green and blue lines represent two distinct EFG distributions, referred  in the text as EFG\textsuperscript{\ce{SMGO1}} and EFG\textsuperscript{\ce{SMGO2}}, respectively.]{\RaggedRight Representative \textit{R(t)} spectra of \ce{SrMnGe2O6} and corresponding fits (left), as well as Fourier transforms (right). Green and blue lines represent two distinct EFG distributions, referred in the text as EFG\textsuperscript{\ce{SMGO1}} and EFG\textsuperscript{\ce{SMGO2}}, respectively.}
\label{Fig:PACSMGO}
\end{figure}

The EFG characterization, after \textsuperscript{111m}Cd implantation, was performed in a wide range of temperatures for both \ce{(Sr/Ca)MnGe2O6} systems. To assess the consistency of the structural features before and after implantation, the pre- and post-implantation X-ray diffractograms are shown in Appendix \ref{Sec:A2}, figures \ref{fig:XRDCMGO} and \ref{fig:XRDSMGO}. This study aimed to determine the location of implanted Cd in the \ce{(Sr/Ca)MnGe2O6} structure and to accurately describe the temperature dependence of the electric field gradient (EFG) in pristine samples. TDPAC measurements on the \ce{SrMnGe2O6} (SMGO) system were conducted at selected temperatures ranging from 13 to 902~K. 
Figure \ref{Fig:PACSMGO} shows a few representative experimental $\gamma - \gamma$ \textit{R(t)} functions obtained for the \ce{SrMnGe2O6} compound along with their corresponding Fourier transforms. Several models were tested in order to numerically obtain the best fit to the experimental \textit{R(t)} spectra before arriving at the presented model, wherein two distinct fractions (two distinct local environments) are evident across the whole temperature range, indicating the probes interacted with two distinct EFG distributions, henceforth referred to as EFG\textsuperscript{\ce{SMGO1}} and EFG\textsuperscript{\ce{SMGO2}}, and represented by the green and blue lines in figure \ref{Fig:PACSMGO}, respectively. 

\begin{figure}[htb]

\centering
\includegraphics[width=0.9 \linewidth]{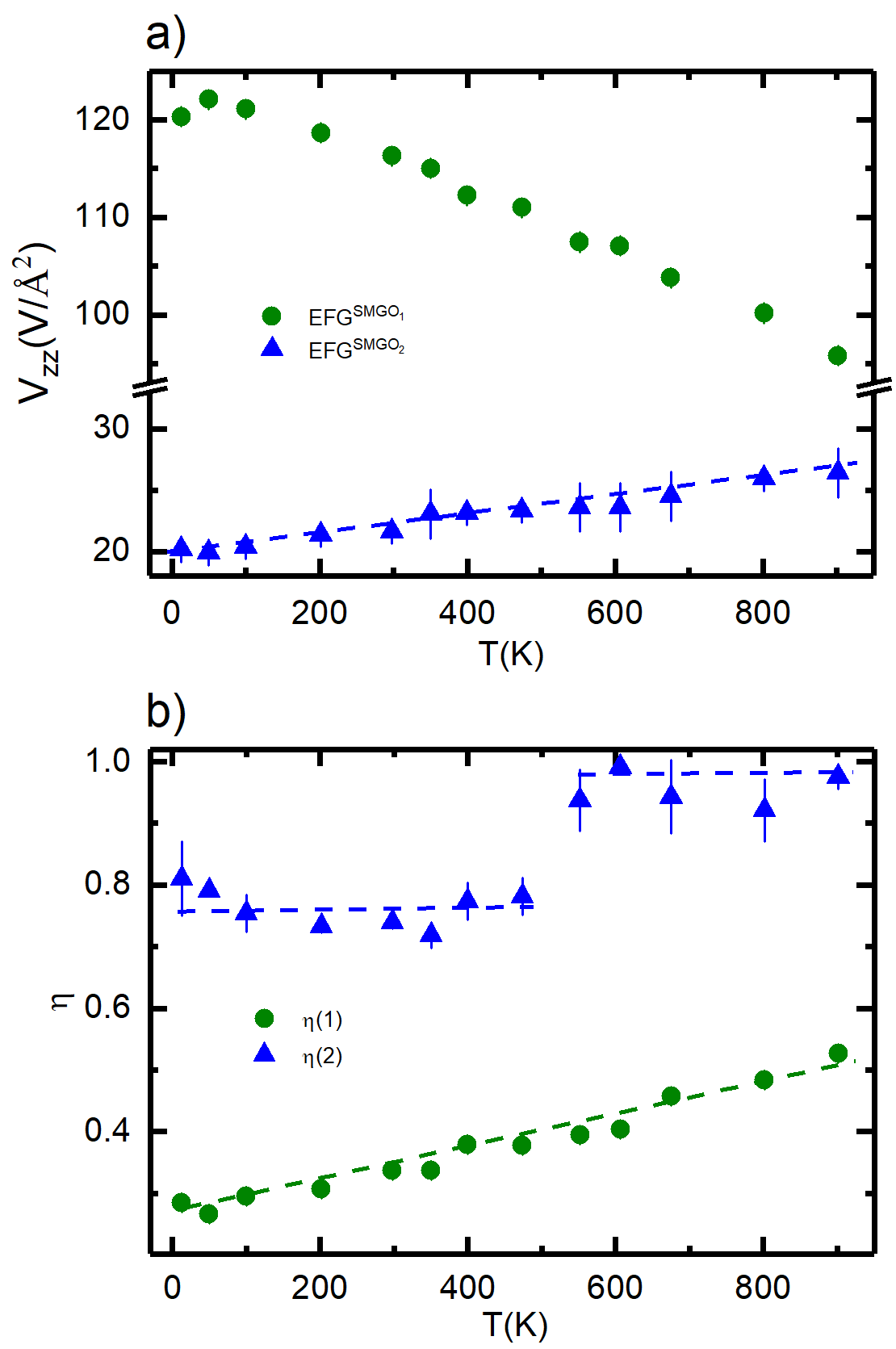}
\caption
{\RaggedRight  Measured EFG parameters at the Cd probe in \ce{SrMnGe2O6}: a) $V_{zz}$ and b) $\eta$. 
The green  circles  correspond to the EFG\textsuperscript{SMGO\textsubscript{1}} fraction and the blue triangles to the  EFG\textsuperscript{SMGO\textsubscript{2}}.  The dashed lines are merely a guide to the eyes. }
\label{Fig:SMGOEFG}
\end{figure}

The temperature behaviour of the experimental \textit{R(t)} fit parameters of SMGO  is displayed in figure \ref{Fig:SMGOEFG}. One can observe, in figure \ref{Fig:SMGOEFG} a), green circles, that the $V_{zz}$ value for the EFG\textsuperscript{\ce{SMGO1}} fraction has a noticeable increase at low temperatures, with increasing temperature, reaching a maximum; whereupon it starts to decrease. On the other hand, the axial asymmetry $\eta(1)$ parameter increases as the temperature increases, as displayed in figure \ref{Fig:SMGOEFG} b) by the green circles. The aforementioned behaviour of the $V_{zz}$ is typically caused by a contribution from low-frequency phonon modes that lead to a linear decrease at high temperatures, as observed, though, the effects of the lattice volume expansion with temperature and structural relaxation around the probe also contribute to this behaviour\cite{Bayer1951,kushida1955influence,kushida1956dependence}.
 Regarding the axial asymmetry parameter $\eta(1)$  at the  EFG\textsuperscript{\ce{SMGO1}} site, it displays an intermediate asymmetry, which increases as temperature raises. 
 
For the EFG\textsuperscript{\ce{SMGO2}} fraction, depicted by the blue triagles 
 in figure \ref{Fig:SMGOEFG} a),  the $V_{zz}$ shows an atypical behaviour of increasing as temperature also increases, though such behaviour has nevertheless been reported in other compounds. Specifically, analogous behaviour was experimentally observed   in the \textit{A}2\textsubscript{1}\textit{am} phase of \ce{Ca3Mn2O7} for which DFT simulations indicated that such behaviour is actually due to structural effects, that prevail over the phonon contributions~\cite{PhysRevB.101.064103}. 
  The asymmetry parameter $\eta(2)$ for the EFG\textsuperscript{\ce{SMGO2}} fraction, as displayed in figure \ref{Fig:SMGOEFG} b) by the blue triangles, 
 on the other hand, remains approximately constant, up to 474~K, whereupon an anomalous increase occurs. It is unclear if such behaviour could be related to structural phenomena, as there are no measurements reported for temperatures above 350~K. 
 Finally, it is worth noting that the probes distribution between both fraction remained fairly constant across the whole temperature range. This strongly suggests that Cd occupies two distinct lattice sites within the crystal and that both the crystallographic structure and Cd positions remain stable up to the highest measured temperature (900 K).

\begin{figure}
\centering
\includegraphics[width=1\linewidth]{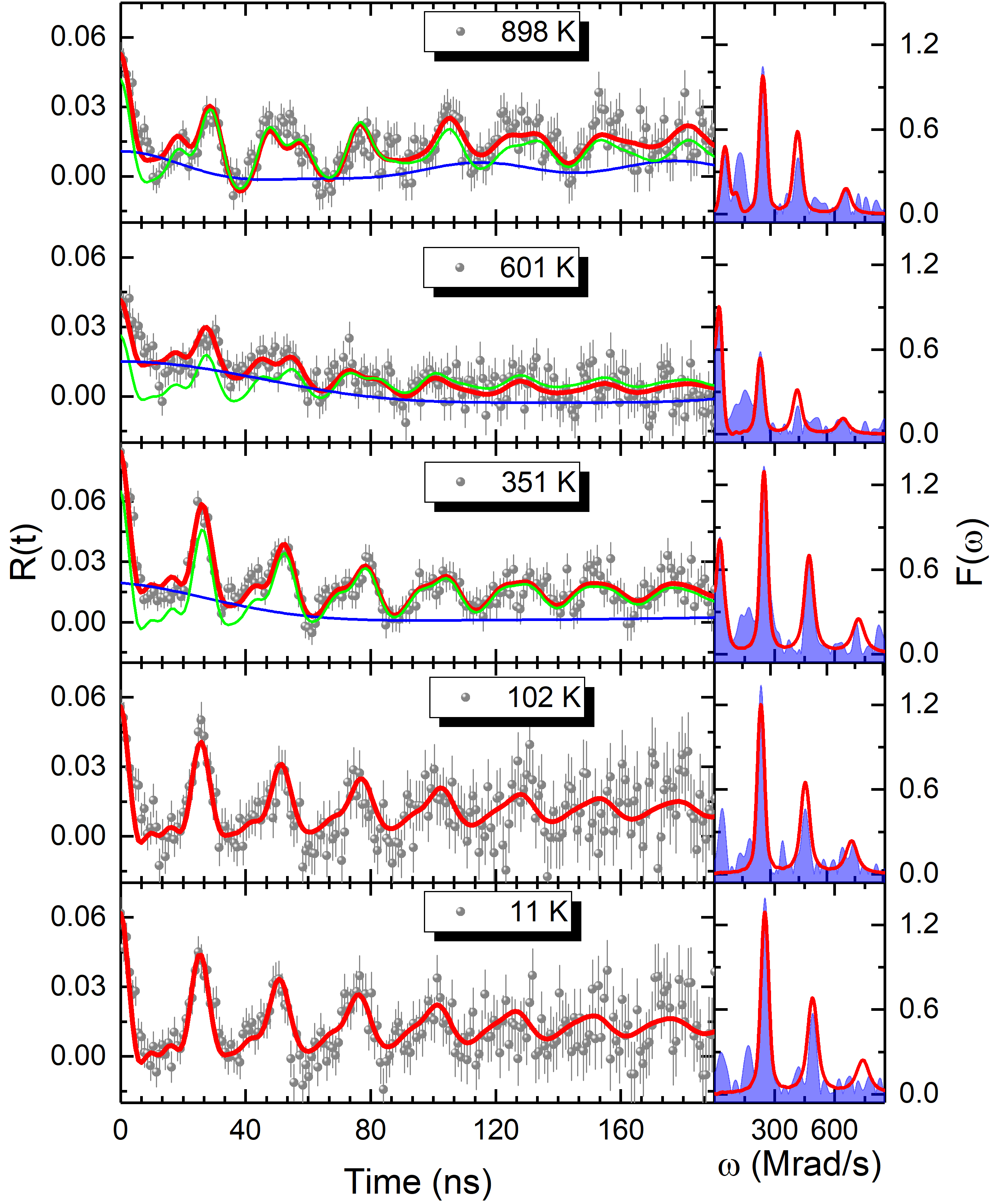}
\caption[\RaggedRight Representative \textit{R(t)} spectra of \ce{CaMnGe2O6} and corresponding fits (left), as well as Fourier transforms (right). Green and blue lines represent two distinct EFG distributions, referred  in the text as EFG\textsuperscript{\ce{CMGO1}} and EFG\textsuperscript{\ce{CMGO2}}, respectively.]{\RaggedRight Representative \textit{R(t)} spectra of \ce{CaMnGe2O6} and corresponding fits (left), as well as Fourier transforms (right). Green and blue lines represent two distinct EFG distributions, referred  in the text as EFG\textsuperscript{\ce{CMGO1}} and EFG\textsuperscript{\ce{CMGO2}}, respectively.}
\label{Fig:PACCMGO}
\end{figure}

With respect to the \ce{CaMnGe2O6} (CMGO) system, the TDPAC measurements were likewise performed at selected temperatures, in the 11-898~K range. Representative fits to the experimental $R(t)$ perturbation functions for CMGO are presented in figure \ref{Fig:PACCMGO}, along with the corresponding Fourier transforms. Similar to how the data from the previous compound was treated,  several models were also tested in order to numerically obtain the best fit to the experimental \textit{R(t)} spectra. Distinctly from \ce{SrMnGe2O6}, wherein two noticeable fractions are present across the whole temperature range, for the \ce{CaMnGe2O6} compound there is only one fraction present from low temperatures, up to 351~K, hereby referred to as EFG\textsuperscript{\ce{CMGO1}} and represented by the green lines in figure \ref{Fig:PACCMGO}. It can be noticed in this figure that a second fraction, labeled EFG\textsuperscript{CMGO2}, appears for temperatures above 351~K, represented by the blue lines. However, it is likely that this second fraction is likewise present at lower temperatures, but owing to the very small observable frequency associated with it, that is not well defined in the time window available in the TDPAC experiments, it is not possible to fit this fraction at lower temperatures. Even at high temperatures the EFG\textsuperscript{CMGO2} fitting parameters have nonetheless an associated high uncertainty.  

\begin{figure}[htb]

\centering
\includegraphics[width=0.9\linewidth]{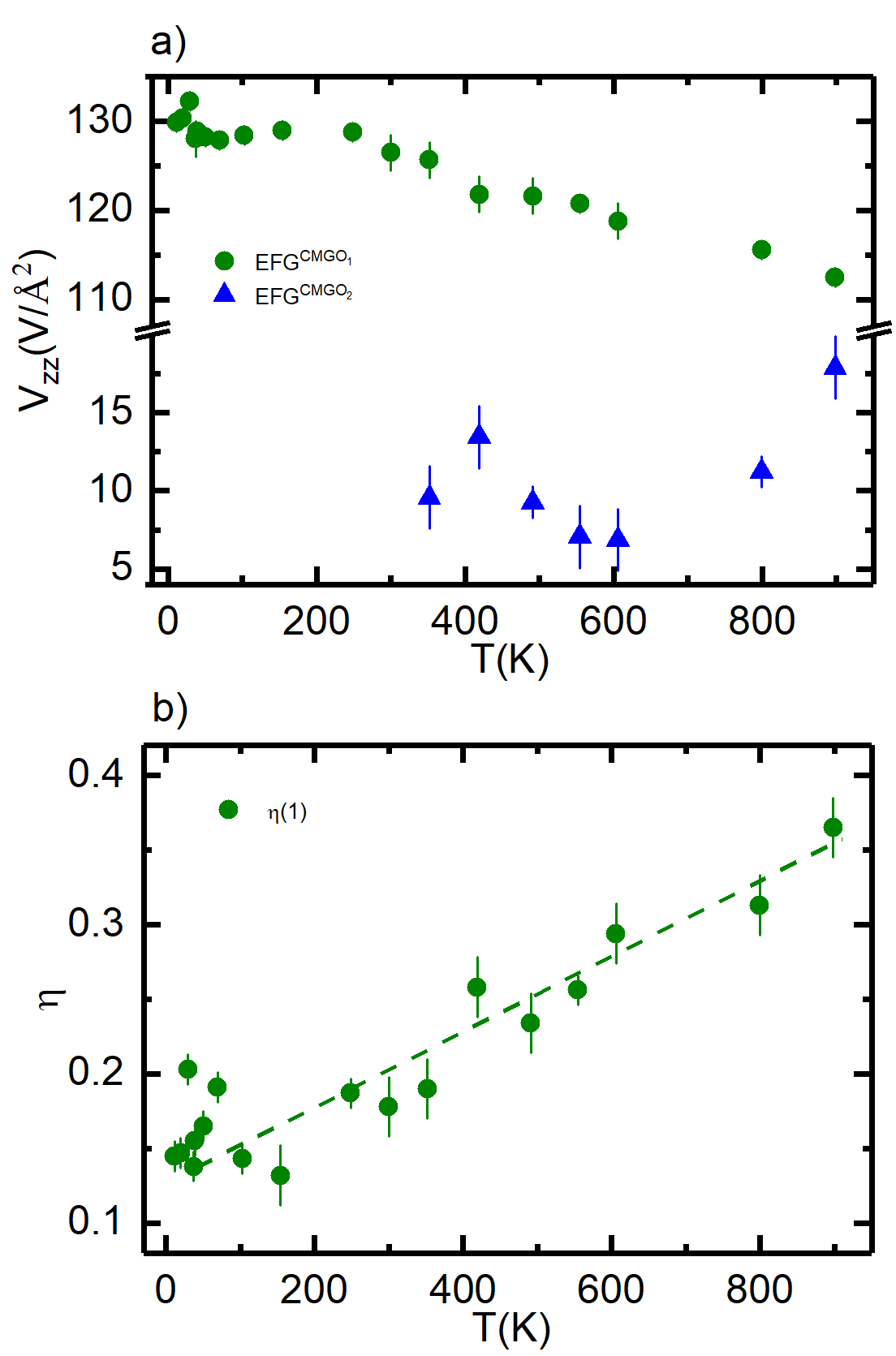}
\caption[\RaggedRight  Measured EFG parameters at the Cd probe in \ce{CaMnGe2O6}: a) $V_{zz}$ and b) $\eta$.  The green  circles  correspond to the EFG\textsuperscript{CMGO\textsubscript{1}} fraction and the blue triangles to the EFG\textsuperscript{CMGO\textsubscript{2}}. The dashed lines are merely a guide to the eyes.] 
{\RaggedRight  Measured EFG parameters at the Cd probe in \ce{CaMnGe2O6}: a) $V_{zz}$ and b) $\eta$.  The green  circles  correspond to the EFG\textsuperscript{CMGO\textsubscript{1}} fraction and the blue triangles to the EFG\textsuperscript{CMGO\textsubscript{2}}.   The dashed lines are merely a guide to the eyes.}  
\label{Fig:CMGOEFG}
\end{figure}

The temperature behaviour of the fit parameters relative to the \ce{CaMnGe2O6} system are shown in figure \ref{Fig:CMGOEFG}. One can observe, in figure \ref{Fig:CMGOEFG} a), green circles,  that the EFG\textsuperscript{\ce{CMGO1}} distribution has a $V_{zz}$ value of similar magnitude to EFG\textsuperscript{SMGO1} (figure \ref{Fig:SMGOEFG} a)), with an analogous decrease as temperature raises. 

Also $\eta(1)$, displayed in figure \ref{Fig:CMGOEFG} b), green circles,  similarly increases with increasing temperature, when compared to SMGO $\eta(1)$ variation (figure \ref{Fig:SMGOEFG} b)). 
Similarly to what was observed for \ce{SrMnGe2O6}, we note that the magnetic hyperfine interactions were not apparent in the fits to the experimental \textit{R(t)} functions.

To aid in the interpretation of these results, experimental data needs to be analyzed in comparison with the EFGs from DFT calculations. These EFGs are summarised in Table \ref{table:EFG_DFT}. 
Comparing the EFG parameters for the experimental EFG\textsuperscript{\ce{SMGO1}} fraction of \ce{SrMnGe2O6}  with those computed for the supercells with Cd substitution at the Sr-site, where the 25\% and 12.5\% Cd dilution was considered (unit-cell and $1 \times 1 \times 2$ supercell, respectively), we may observe a very good agreement between $V_{zz}$ and $\eta$. From the TDPAC measurements we obtain for EFG\textsuperscript{\ce{SMGO1}}the values of $V_{zz}$=120 V/\AA$^2$ and $\eta$=0.28; for the Cd probe nucleus for \ce{Sr_{0.75}Cd_{0.25}MnGe2O6} they are $V_{zz}$=103 V/\AA$^2$  and $\eta$=0.14 and for \ce{Sr_{0.875}Cd_{0.125}MnGe2O6} these result as $V_{zz}$=108 V/\AA$^2$ and $\eta$=0.14. Likewise, good agreement is also observed for the EFG\textsuperscript{\ce{SMGO2}} fraction of \ce{SrMnGe2O6} for the case of the Cd substitution at the Mn-site.
We obtain for the EFG\textsuperscript{\ce{SMGO2}} fraction the values of $V_{zz}$=20 V/\AA$^2$ and $\eta$=0.81. From the DFT results we get $V_{zz}$=22 V/\AA$^2$ and $\eta$=0.67, and $V_{zz}$=21 V/\AA$^2$ and $\eta$=0.72, for the \ce{SrMn_{0.75}Cd_{0.25}Ge2O6} and \ce{SrMn_{0.875}Cd_{0.125}Ge2O6} cells, respectively. 
Based on the obtained theoretical results, we may conclude that the supercells simulate quite well the experimental data when the Cd probe is considered, with the EFGs being in good agreement to the EFG\textsuperscript{\ce{SMGO1}} and EFG\textsuperscript{\ce{SMGO2}} experimental fractions, which correspond to the Cd probe replacing the Sr-sites and the Mn-sites, respectively, in the crystalline structure. 

 Similarly, this same correspondence seems to hold true for the \ce{CaMnGe2O6} compound. 
For the EFG\textsuperscript{\ce{CMGO1}} experimental fraction, we observe the values of $V_{zz}$=130 V/\AA$^2$ and $\eta$=0.15, and for the DFT data we obtain values of $V_{zz}$=108 (109)  V/\AA$^2$ and $\eta$=0.01 (0.02) for  Cd concentration of 1/4 (1/8)  in the Ca atom position.  
The obtained theoretical EFGs are slightly lower than those of the experimental fraction; such is mostly evident for the asymmetry parameter, $\eta$. As for the EFG\textsuperscript{\ce{CMGO2}} fraction from the TDPAC experiments, the EFG  result in $V_{zz}$=10 V/\AA$^2$ and an indeterminate $\eta$, whereas for the DFT diluted supercell of \ce{CaMn_{0.825}Cd_{0.125}Ge2O6} we have $V_{zz}$=12 V/\AA$^2$ and $\eta$=0.21, similar to those obtained for \ce{CaMn_{0.75}Cd_{0.25}Ge2O6}. 
It should be noted that in the cases where Cd was placed at a Mn site, the systems were treated as ferromagnetic, for the sake of computational simplicity. An antiferromagnetic calculation for the \ce{CaMn_{0.875}Cd_{0.125}Ge2O6} system was nevertheless performed to check how much of an effect the magnetic ordering might have on the computed EFG. Very similar results were obtained in both the ferromagnetic and antiferromagnetic calculations, with $V_{zz}$=12 V/\AA$^2$ and $\eta$=0.21, in the former case, 
and $V_{zz}$=13 V/\AA$^2$ and $\eta$=0.37, in the latter, 
thus showing that this approximation is reliable.  

These results indicate that first-principles calculations combined with TDPAC spectroscopy provide a valuable tool to explain the observed two distinct local environments, indicating that Cd probes can in fact be distributed between the Sr- or Ca-sites and the Mn-sites in the \ce{SrMnGe2O6} and \ce{CaMnGe2O6} basic compounds, respectively.

\subsection{X-ray diffraction}
Based on the comparison of TDPAC measurements with DFT calculations, which indicated that Cd probes occupy either the \textit{A} or the Mn  cation sites in pristine systems,
\begin{figure}[h]
    
    \centering
    \includegraphics[width=0.42\textwidth]{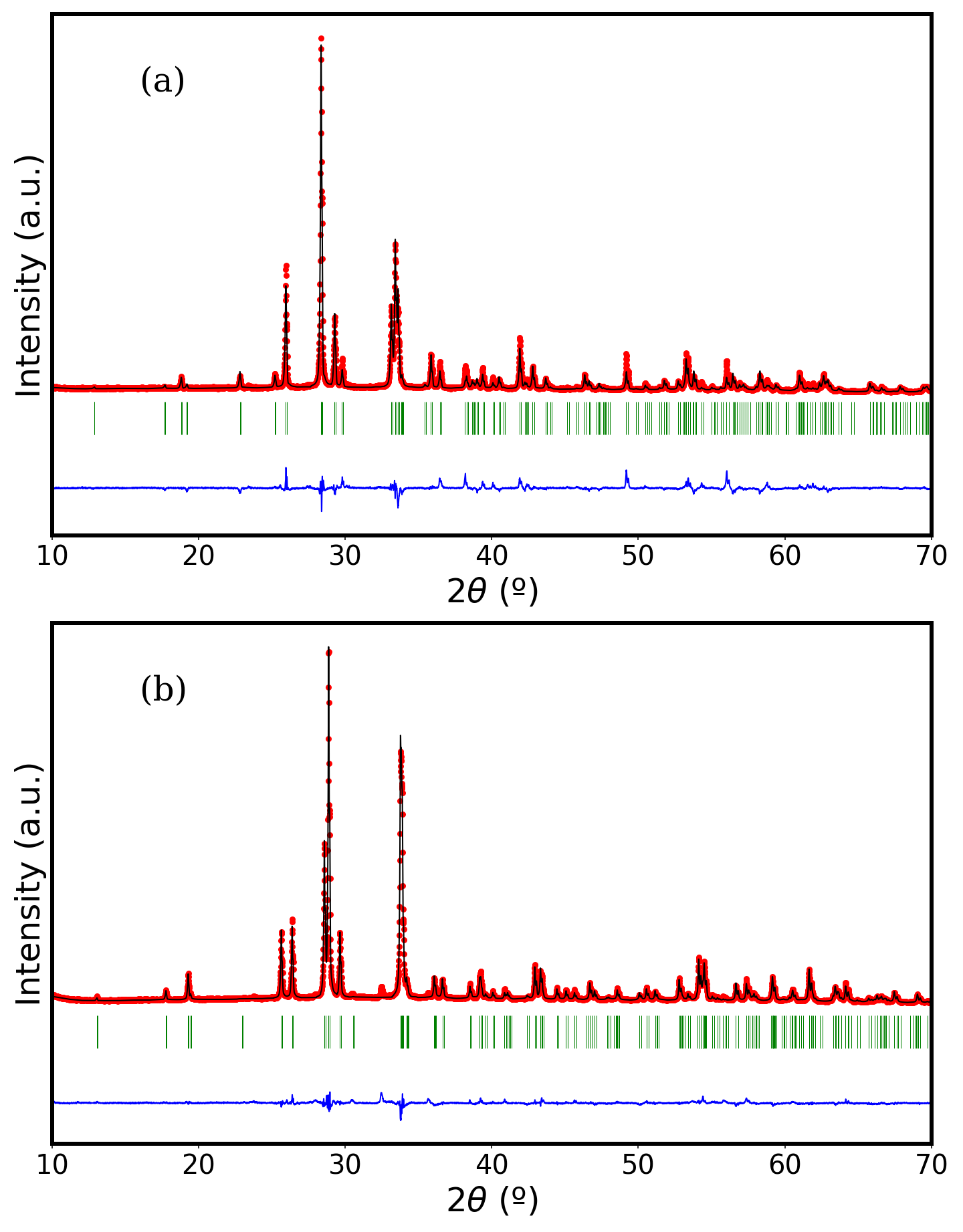}
    \caption{\RaggedRight Experimental XRD pattern (red dots), calculated XRD pattern based on the structure obtained from Rietveld refinement (black line) and difference between the two curves (blue line) of the: (a) \ce{Sr_{0.875}Cd_{0.125}MnGe2O6} and (b) \ce{Ca_{0.875}Cd_{0.125}MnGe2O6}. Green tick marks indicate the position of the Bragg reflections for the $C2/c$ space group.}
    \label{fig:XRDSr0.875&Ca.875}
\end{figure}
and considering that \textit{ab initio} calculations suggested Cd doping may impact the band gaps of these systems, we attempted to synthesize doped samples of \ce{CaMnGe2O6} and \ce{SrMnGe2O6} at the same concentrations as were considered for the DFT calculations.

In order to gauge the phase purity of the samples, XRD measurements were performed, along with a Reitveld refinement that was done to obtain the lattice parameters and atomic positions of the new compounds. 
The structural refinement was conducted using the room temperature XRD data, with the previously reported models for the parent compounds \cite{Ding_JMCC2016,Ding_PRB2016}.
The X-ray data for the \ce{Sr_{0.875}Cd_{0.125}MnGe2O6} and the \ce{Ca_{0.875}Cd_{0.125}MnGe2O6} and their respective Rietveld refinements are shown in figure \ref{fig:XRDSr0.875&Ca.875} whilst those for the remainder of the systems are in Appendix \ref{App:Xray}.

\begin{figure}[h]
    
    \centering
    \includegraphics[width=0.42\textwidth,height=9cm]{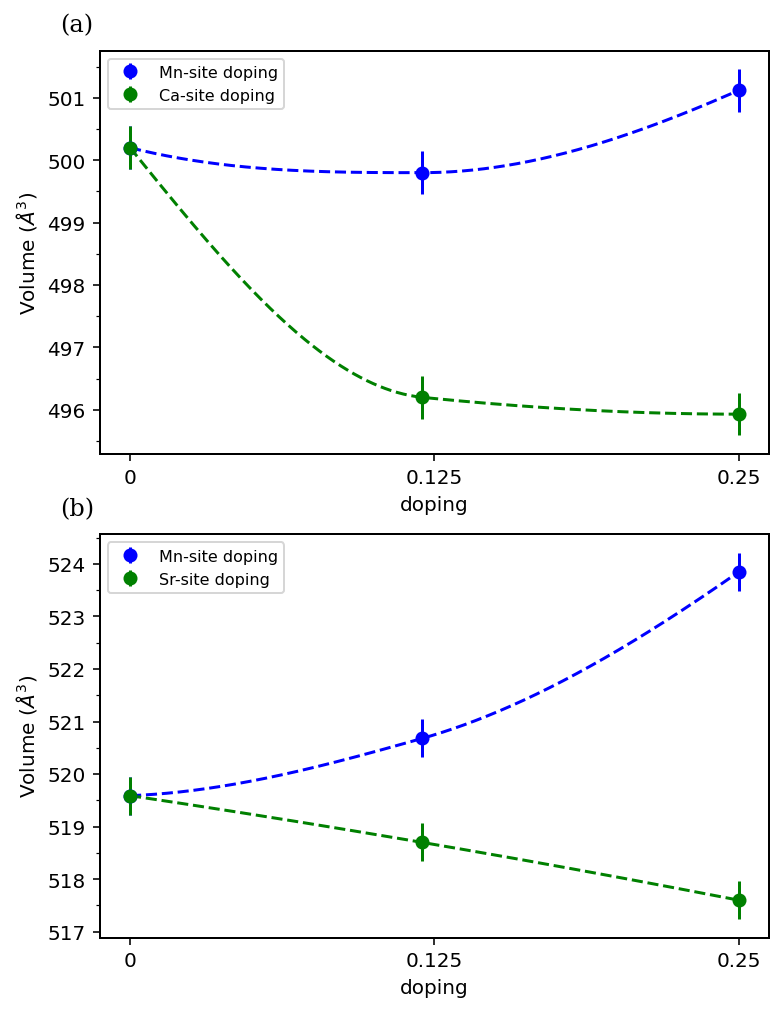}
    \caption{\RaggedRight Volume as a function of doping for: (a) \ce{CaMnGe2O6} (b) \ce{SrMnGe2O6}.}
    \label{fig:Volumedoping}
\end{figure}

The X-ray measurements showed that all doped samples crystallized in the $C2/c$ space group, the same as the parent compounds, along with some minor impurity phases (the remaining experimental XRD patterns are shown in Appendix \ref{App:Xray}). The lattice parameters and atomic positions are given in \cref{table:Parameters,table:PositionsSr0.875,table:PositionsSr0.75,table:PositionsSrMn0.875,table:PositionsSrMn0.75,table:PositionsCa0.875,table:PositionsCa0.75,table:PositionsCaMn0.875,table:PositionsCaMn0.75}. 
The volume dependence on doping for the systems is represented in figure \ref{fig:Volumedoping}. A slight decrease in cell volume (<0.9\%) is observed by doping the Ca/Sr site, while a small increase (0.85\% ) is noticed when alloying at the manganese site. These trends align with expectations based on the atomic radii of the dopant elements. 

It can be observed that the doping at the \textit{A} site led to a much more significant reduction in cell volume in the \ce{CaMnGe2O6} system, relative to \ce{SrMnGe2O6}. In contrast, doping at the Mn site had a more pronounced effect in increasing the volume of \ce{SrMnGe2O6} as compared to \ce{CaMnGe2O6}. These observations may point towards the Cd dopants preferring the Ca over the Mn site in \ce{CaMnGe2O6}, whereas in \ce{SrMnG2O6}, the Sr site is more favourable for doping. Although according to structural data from references \onlinecite{Ding_PRB2016,PhysRevB.101.235109} the \ce{MnO6} octahedra in these systems have similar volumes, the octahedra are observed to be more distorted in \ce{CaMnGe2O6} than in \ce{SrMnGe2O6}, and may therefore be less favourable of an environment for Cd. This is in agreement with the results from the TDPAC measurements, where the fraction identified as occupying the Ca  site is well-defined and present across the whole temperature range, whereas the fraction that was identified as occupying the Mn site is only found at higher temperatures and is overall less clearly defined than the first one, likewise suggesting that the Ca site is less favourable for occupation by Cd atoms than the Mn site.

\FloatBarrier

\subsection{Reflectometry}

To investigate the optical properties of synthesized compounds, we recorded the diffuse reflectance spectra in the UV-visible region 200–900 nm at room temperature. 

\begin{figure}[h]  
    \centering
    \includegraphics[width=0.42\textwidth]{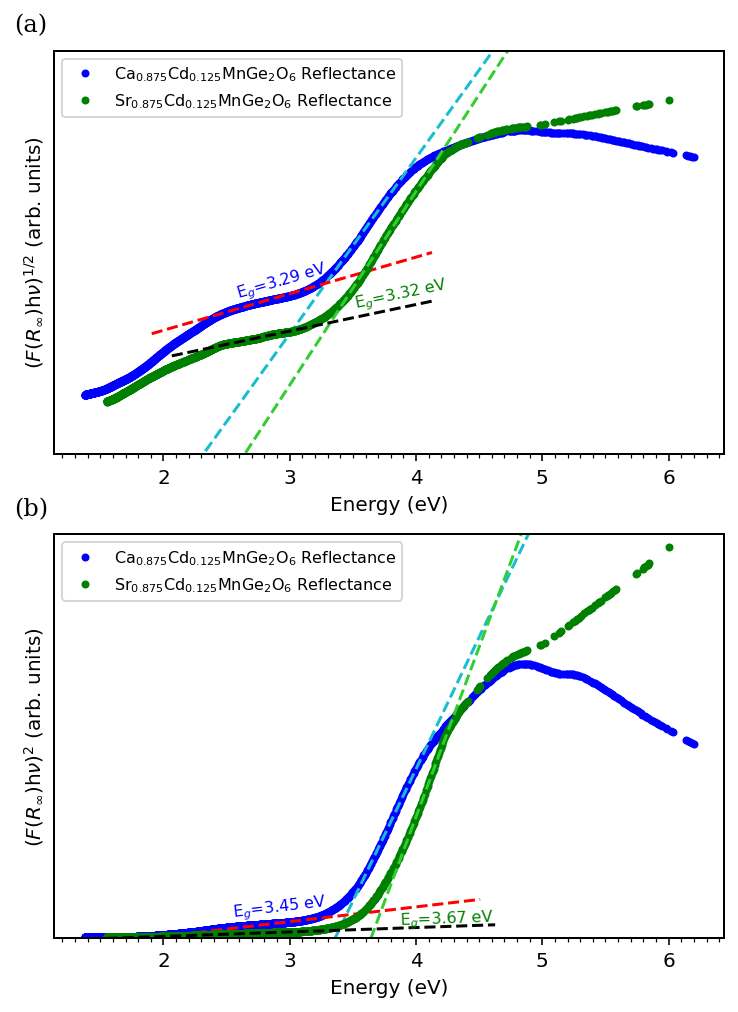}
    \caption{\RaggedRight Tauc's plot of (a) indirect and (b) direct bandgap $E_g$ for the \ce{Ca_{0.875}Cd_{0.125}MnGe2O6} and \ce{Sr_{0.875}Cd_{0.125}MnGe2O6} compounds. Dashed lines represent linear fits.}
     \label{fig:Reflectance}
\end{figure}

Figure \ref{fig:Reflectance} illustrates representative Tauc plots of the Kubelka–Munk function for direct and indirect allowed transitions obtained from the diffuse reflectance spectra. The experimental data shows that the samples start to absorb for $\lambda$ < 420 nm (> 2.7 eV), usually corresponding to the threshold of indirect optical band gap, leading to prompt e-/h+ recombination or only reaching the conduction band by adding extra energy from the crystalline (thermal) phonons.
The absorption reaches maximum at $\lambda$ $\sim$ 340 nm ($\sim$ 3.6 eV) corresponding to the effective band gap or the edge of direct optical band gap. These photons energy can drive directly the photo-excited electrons to the conduction band (minimizing recombination) until filling the limited density of states available. Higher energy conduction bands (with limited DOS) are available between 4 and 5 eV. Above 5 eV further transport mechanisms can be activated.
\begin{figure}
   
    \centering
    \includegraphics[width=0.48\textwidth]{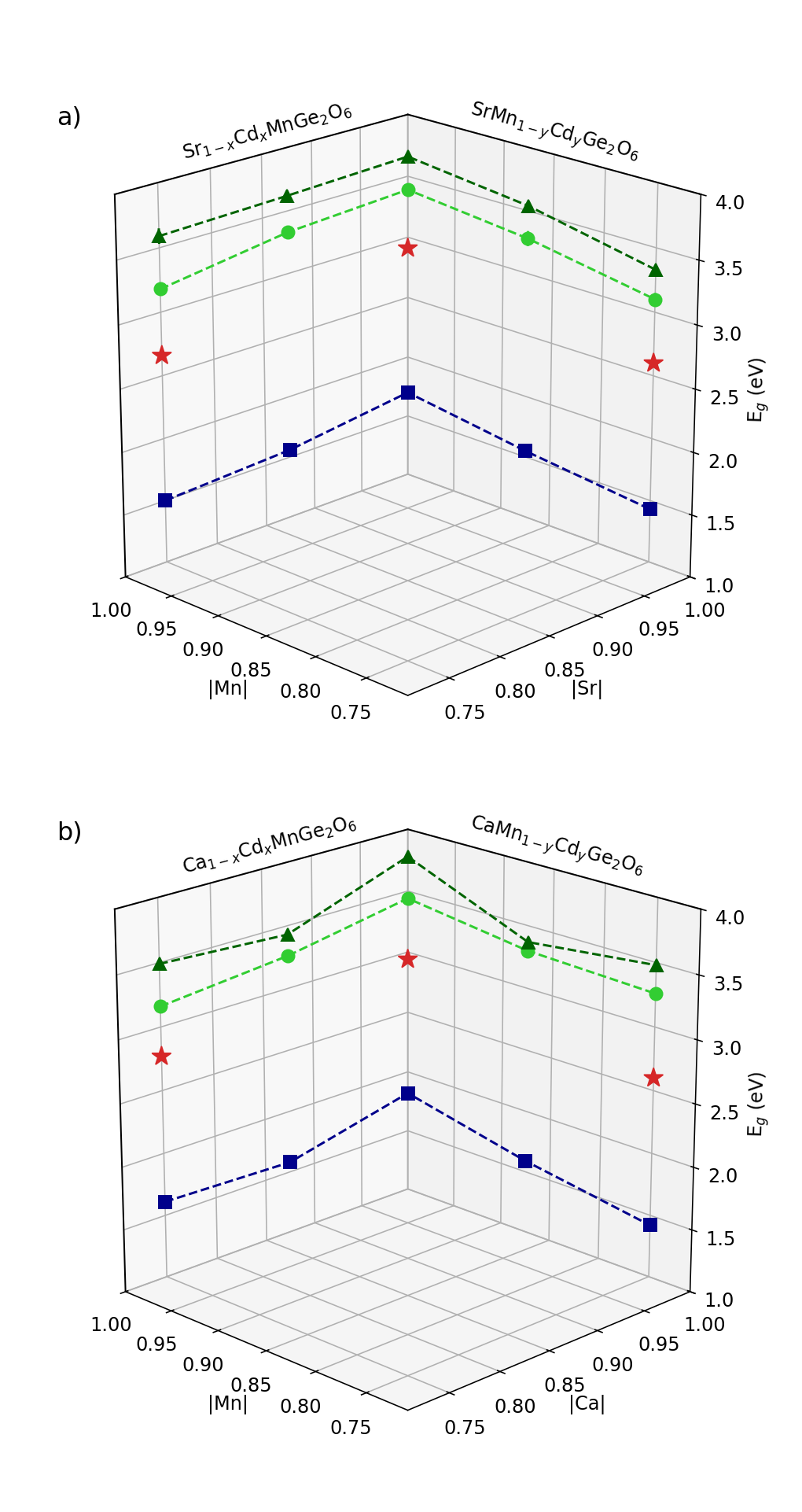}
    \caption{\RaggedRight Experimental direct (dark green triangles) and indirect (green circles) bandgaps, along with the values 
    obtained with DFT+U  approach (blue squares) and HSE06 hybrid functional (red stars) as a function of Cd concentration  $x$ for: (a) \ce{SrMnGe2O6} doped with \ce{Cd} at the Sr and Mn sites (b) \ce{CaMnGe2O6} doped with \ce{Cd} at the Ca and Mn sites.}
     \label{fig:Gaps}
\end{figure}
After determining the intersection in energy of the linear fit to the fundamental absorption edge with the base line defined by a linear fit to the weak slope below the fundamental absorption, one obtains the values of the energy gap transition. The experimental change in the band gap width upon alloying with Cd at Ca/Sr or Mn sites, along with those obtained from  PBE+$U$ calculations, as well as calculations utilizing the HSE06 hybrid functional are presented in Table \ref{table:gaps} and shown in figure \ref{fig:Gaps}.
The results showed that alloying with Cd in these systems lead to a reduction in band gap width, as was predicted by the DFT calculations. Moreover, with the exception of doping at the Ca$^{2+}$ site in \ce{CaMnGe2O6} where the band gap remained close to constant with increased Cd content, a further reduction was observed when increasing the Cd amount from 12.5\% to 25\%. Whereas the parent systems have band gaps that lie beyond the upper limit of the visible light spectrum, alloying with Cd at an amount of 25\%, the band gap is lowered in value to lie within the spectrum of visible light.

\FloatBarrier
\section{Conclusions}
We have undertaken DFT calculations to obtain the structural and electronic properties of the \ce{\textit{A}MnGe2O6} clinopyroxene systems, with $\textit{A}=\rm {Be, Mg, Ca, Sr}$. We found, through enthalpy of formation calculations, that the Be- and Mg-based systems
are not thermodynamically stable against dissociation into the stable constituent oxides \ce{BeGeO3} or \ce{MgGeO3} and \ce{MnGeO3}. On the other hand, the Ca- and Sr-based systems are thermodynamically stable against dissociation and, accordingly,  have been realized experimentally. From the PDOS, we found that the strongly correlated Mn $3d$-states form the narrow band close to the VBM together with the O $p$-states.  The CBM is composed mainly of low density hybridized $p$-states of O and Ge and, for higher energies, the density of states significantly increases
forming a narrow resonant band that is mainly composed of the anti-bound unnocupied Mn $3d$-states. The theoretical electronic band gaps of \ce{CaMnGe2O6} and \ce{SrMnGe2O6} compounds, computed using the PBE+$U$ approximation were found to be significantly underestimated by comparison to the experimental optical gaps obtained through reflectometry, whereas calculations using the HSE06 hybrid functional were able to closely replicate the experimental results. The calculated effective masses of the charge carriers of the two studied pristine compounds show considerable differences, which implies that the separation of the electron–hole pairs can be enhanced, thus improving the photoelectric efficiency. By considering the Cd dopants, we do not observe much variation on the value of the charge carrier masses, whereas a decrease of the band-gap widths is observed.

Through the combination of theoretical {\it{ab-initio}} calculations and TDPAC results, we could infer that the \ce{Cd} probe in the \ce{\textit{A}MnGe2O6} clinopyroxene compounds, with $\textit{A}=\rm {Ca, Sr}$, can replace 
the ion located at the $A$-site as well as the Mn ion.
Subsequently, the new germanate clinopyroxene Cd alloys were successfully synthesized and experimentally characterized. The diffuse reflectance measurements show the possibility of tuning the band gap through Cd doping and the absorption for UV photons, with the possibility of tuning the absorption towards the upper limits of the visible spectrum through Cd doping, highlighting its potential applications in various fields.

\begin{acknowledgments}

The authors acknowledge the support of the technical teams at ISOLDE for their exceptional work in delivering high-quality beams for the presented TDPAC measurements. The authors also acknowledge project NECL under NORTE-01-0145-FEDER-022096, and FCT projects UIDP/04968/2020 (doi.org/10.54499/UIDP/04968/2020), UIDB/04968/2020 (doi.org/10.54499/UIDB/04968/2020), LaP-MET LA/P/0095/2020, POCI-01-0145-FEDER-029454, POCI-01-0145-FEDER-032527, CERN/FIS-TEC/0003/2021 (https://doi.org/10.54499/CERN/FIS-
TEC/0003/2021) and 2024.00223.CERN (https://doi.org/10.54499/2024.00223.CERN). Also, BMBF through grants 05K16PGA and 05K22PGA, EU Horizon Europe Framework research and innovation programme under grant agreement no. 101057511 (EURO-LABS) for supporting IS679 ISOLDE-CERN experiment, and PRACE project with reference 2021240118, with access to the Irene Skylake computer. RPM acknowledges support from the Project HPC-EUROPA3 (INFRAIA-2016-1-730897), with the support of the EC Research Innovation Action under the H2020 Programme, with reference HPC171030W and from FCT through the PhD studentship with reference 2020.08546.BD; in particular, the support of Dr. A. Stroppa and the hospitality provided by CNR-SPIN at the Department of Physical and Chemical Science of the University of L'Aquila (Italy) during the visit from 15/10/2021 
to 20/12/2021 as well as computer resources and technical support provided by CINECA. RPM also acknowledges computer resources and technical support from Minho Advanced Computing Center through project CPCA/A1/460622/2021, financed by FCT. LVCA (Project 314884/2021-1) and HMP (Project 308438/2022-1) acknowledge funding from CNPq and support from FAPESP (Project 2022/10095-8). AMLL acknowledge the FCT 2021.04084.CEECIND (doi.org/\-10.54499/\-2021.04084.CEECIND/\-CP1655/\-CT0002) and ELdS   the 2022.00082.CEECIND (doi.org/\-10.54499/\-2022.00082.CEECIND/\-CP1719/\-CT0001) grants. ELdS further acknowledges the High Performance Computing Chair - a R\&D infrastructure (based at the University of Évora; PI: M. Avillez), endorsed by Hewlett Packard Enterprise, and involving a consortium of higher education institutions, research centers, enterprises, and public/private organizations. 

\end{acknowledgments}

\appendix
\label{Sec:Appendix}

\renewcommand\thefigure{\thesection\arabic{figure}} 
\renewcommand\thetable{\thesection\arabic{table}}   
\renewcommand\thesection{\Alph{section}}
\renewcommand\thesubsection{\thesection.\arabic{subsection}}

\section{Formation Energies}
\label{Sec:A1}
\setcounter{table}{0}
\setcounter{figure}{0}

The formation energies for the studied oxides of the  \ce{\textit{A}MnGe2O6} (\textit{A} = Be, Mg, Ca, Sr) clinopyroxene series were calculated according to the following equation:
\begin{equation}
\label{eq:formation}
 \begin{aligned}
E_{form} = E_{\ce{\textit{A}MnGe2O6}} - E_{\ce{\textit{A}GeO3}} - E_{\ce{MnGeO3}},
 \end{aligned}
 \end{equation}

\noindent
where $E_{\ce{\textit{A}MnGe2O6}}$, $E_{\ce{\textit{A}GeO3}}$, and $E_{\ce{MnGeO3}}$ are the total electronic energies obtained from DFT+$U$ calculations, per formula unit (f.u.), which are summarized in  table \ref{table:energies}.

\begin{table}[!h]
\vspace{0.1cm}
\centering
\renewcommand{\arraystretch}{1.8} 
\setlength{\tabcolsep}{9pt} 
 \caption{\RaggedRight Electronic total energies of the  \ce{\textit{A}MnGe2O6} clinopyroxene compounds (\textit{A} = Be, Mg, Ca, Sr)  and  of their respective  constituent  \ce{\textit{A}GeO3} and \ce{MnGeO3} oxides.}
\begin{tabular}{|lc|}
\hline \hline
~~~~~System & Energy (eV/f.u.) \\
\hline \hline
~~SrMnGe$_2$O$_6$  &  -19433.62 \\
\hline
 ~~CaMnGe$_2$O$_6$ &  -17709.36 \\
\hline
~~MgMnGe$_2$O$_6$  &  -17815.41 \\
\hline
~~BeMnGe$_2$O$_6$  &  -15984.00 \\
\hline
~~SrGeO$_3$        &  -9540.40  \\
\hline
~~CaGeO$_3$        &  -7816.06 \\
\hline
~~MgGeO$_3$        &  -7922.42 \\
\hline
~~BeGeO$_3$        &  -6091.01 \\
\hline
~~MnGeO$_3$        &  -9893.11 \\
\hline \hline
\end{tabular}
\label{table:energies}
\end{table}

\section{X-ray Diffraction}
\label{App:Xray}
\setcounter{table}{0}
\setcounter{figure}{0}
The XRD measurements for the pristine \ce{CaMnGe2O6} and \ce{SrMnGe2O6} systems, before and after implantation with \textsuperscript{111m}Cd probes is presented in figures \ref{fig:XRDCMGO} and \ref{fig:XRDSMGO}.
The X-ray data for \ce{Ca_{1-x}Cd_{x}MnGe2O6} and \ce{CaCd_{x}Mn_{1-x}Ge2O6}   ($x=0.125,0.25$) are presented, respectively, in figures \ref{fig:XRDCa0.875} to \ref{fig:XRDCaMn0.75} and for 
\ce{Sr_{1-x}Cd_{x}MnGe2O6} and \ce{SrCd_{x}Mn_{1-x}Ge2O6} ($x=0.125,0.25$)  the X-ray data are shown in figures 
\ref{fig:XRDSr0.875} to \ref{fig:XRDSrMn0.75}, along with their respective Rietveld refinements.  
All refined structural parameters are shown in tables \ref{table:PositionsCa0.875} to \ref{table:PositionsSrMn0.75}.
Table \ref{table:Parameters} displays the refined lattice parameters for all Cd-related compounds. 


\begin{figure*}[h]

    \centering
    \includegraphics[width=1\textwidth,height=20cm]{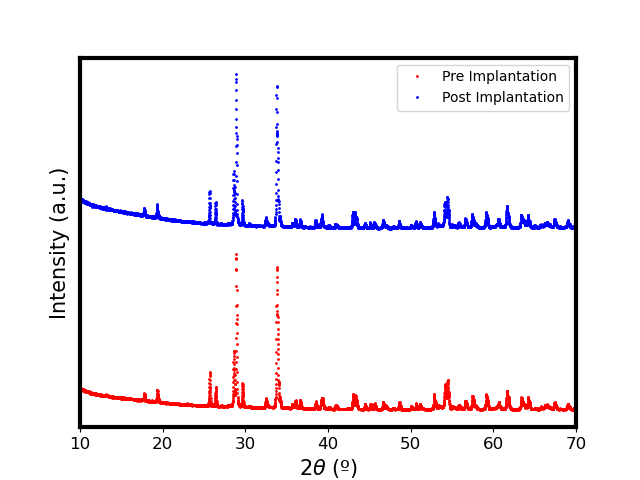}
    \caption{\RaggedRight Experimental XRD pattern for \ce{CaMnGe2O6}, before (red dots) and after (blue dots) implantation with Cd.}
     \label{fig:XRDCMGO}
\end{figure*}

\begin{figure*}[h]

    \centering
    \includegraphics[width=1\textwidth,height=20cm]{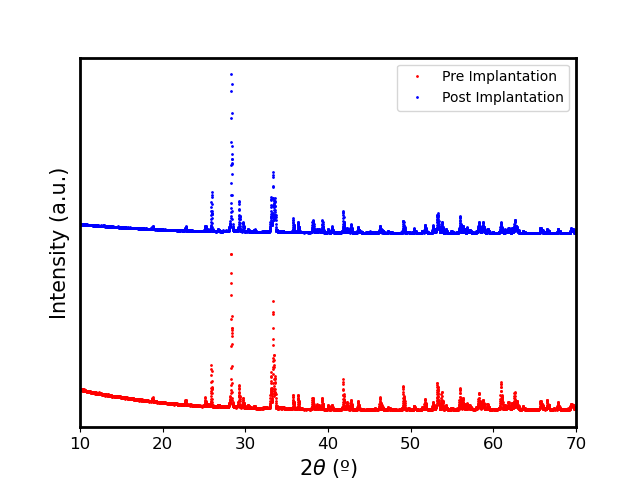}
    \caption{\RaggedRight Experimental XRD pattern for \ce{SrMnGe2O6}, before (red dots) and after (blue dots) implantation with Cd.}
     \label{fig:XRDSMGO}
\end{figure*}

\begin{figure*}[h]

    \centering
    \includegraphics[width=1\textwidth,height=10cm]{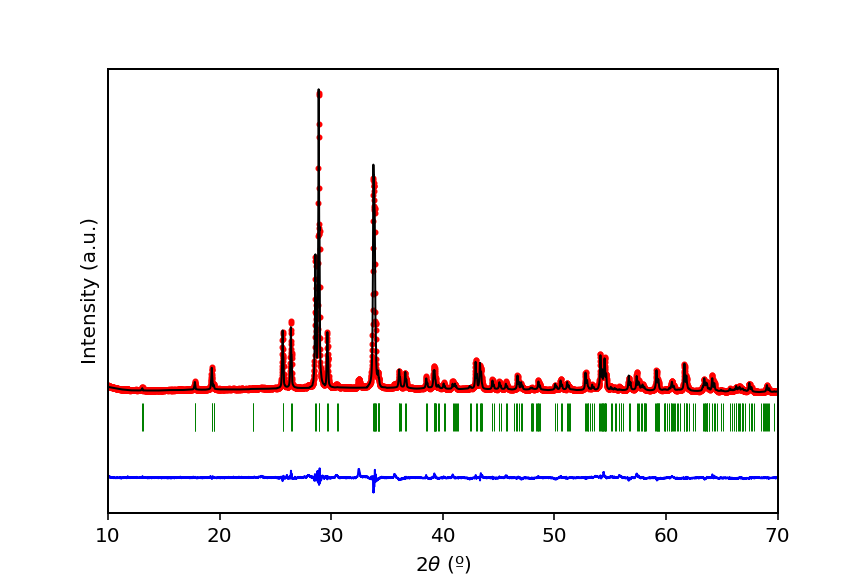}
    \caption{\RaggedRight Experimental XRD pattern (red dots), calculated XRD pattern based on the structure obtained from Rietveld refinement (black line) and the two curves difference (blue line) of the \ce{Ca_{0.875}Cd_{0.125}MnGe2O6} compound. Green tick marks indicate the position of the Bragg reflections for the $C2/c$ space group.}
     \label{fig:XRDCa0.875}
\end{figure*}

\begin{figure*}[h]
    \centering
    \includegraphics[width=1\textwidth,height=10cm]{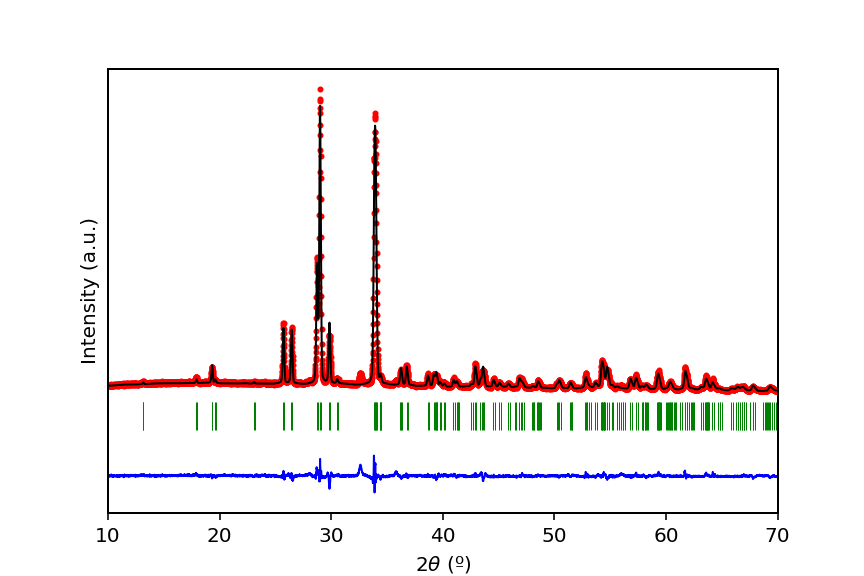} 
    \caption{\RaggedRight Experimental XRD pattern (red dots), calculated XRD pattern based on the structure obtained from Rietveld refinement (black line) and the two curves difference (blue line) of the \ce{Ca_{0.75}Cd_{0.25}MnGe2O6} compound. Green tick marks indicate the position of the Bragg reflections for the $C2/c$ space group.}
    \label{fig:XRDCa0.75}
\end{figure*}

\begin{figure*}[h]
    \centering
    \includegraphics[width=1\textwidth,height=10cm]{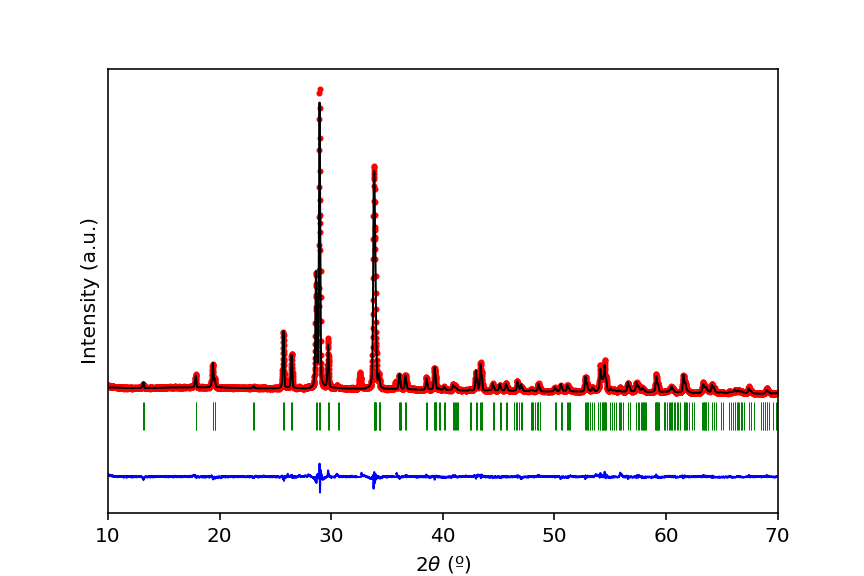}
    \caption{\RaggedRight Experimental XRD pattern (red dots), calculated XRD pattern based on the structure obtained from Rietveld refinement (black line) and the two curves difference (blue line) of the \ce{CaMn_{0.875}Cd_{0.125}Ge2O6} compound. Green tick marks indicate the position of the Bragg reflections for the $C2/c$ space group.}
    \label{fig:XRDCaMn0.875}
\end{figure*}

\begin{figure*}[h]    
    \centering
   \includegraphics[width=1\textwidth,height=10cm]{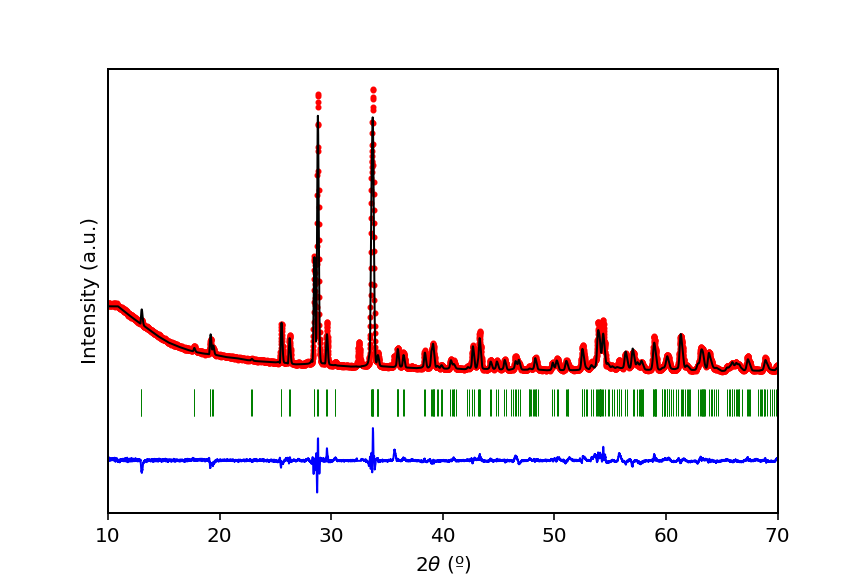}
    \caption{\RaggedRight Experimental XRD pattern (red dots), calculated XRD pattern based on the structure obtained from Rietveld refinement (black line) and difference between the two curves (blue line) of the \ce{CaMn_{0.75}Cd_{0.25}Ge2O6} compound. Green tick marks indicate the position of the Bragg reflections for the $C2/c$ space group.}
        \label{fig:XRDCaMn0.75}
\end{figure*}

\begin{figure*}[h]
     \centering
    \includegraphics[width=1\textwidth,height=10cm]{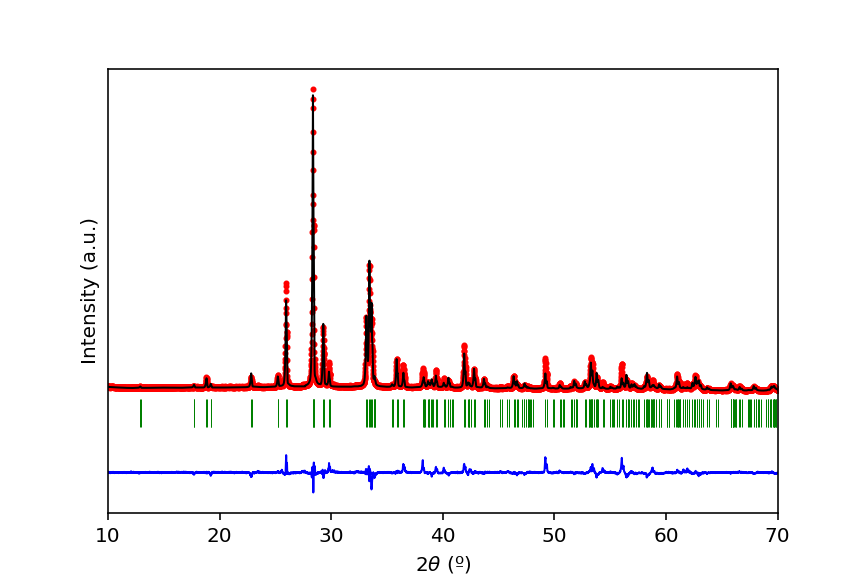}
    \caption{\RaggedRight Experimental XRD pattern (red dots), calculated XRD pattern based on the structure obtained from Rietveld refinement (black line) and difference between the two curves (blue line) of the \ce{Sr_{0.875}Cd_{0.125}MnGe2O6} compound. Green tick marks indicate the position of the Bragg reflections for the $C2/c$ space group.}
    \label{fig:XRDSr0.875}
\end{figure*}

\begin{figure*}[h]   
    \centering
    \includegraphics[width=1\textwidth,height=10cm]{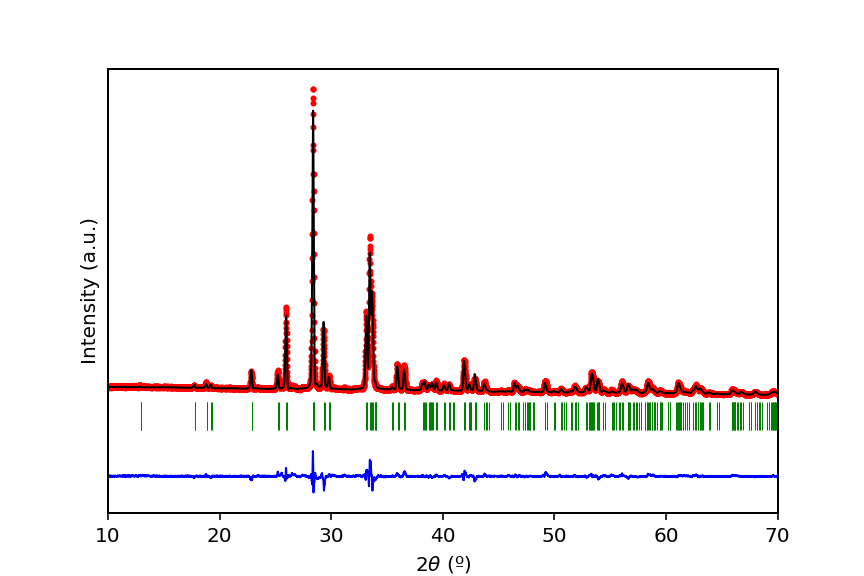}
    \caption{\RaggedRight Experimental XRD pattern (red dots), calculated XRD pattern based on the structure obtained from Rietveld refinement (black line) and difference between the two curves (blue line) of the \ce{Sr_{0.75}Cd_{0.25}MnGe2O6} compound. Green tick marks indicate the position of the Bragg reflections for the $C2/c$ space group.}
    \label{fig:XRDSr0.75}
\end{figure*}

\begin{figure*}[h]
    \centering
   \includegraphics[width=1\textwidth,height=10cm]{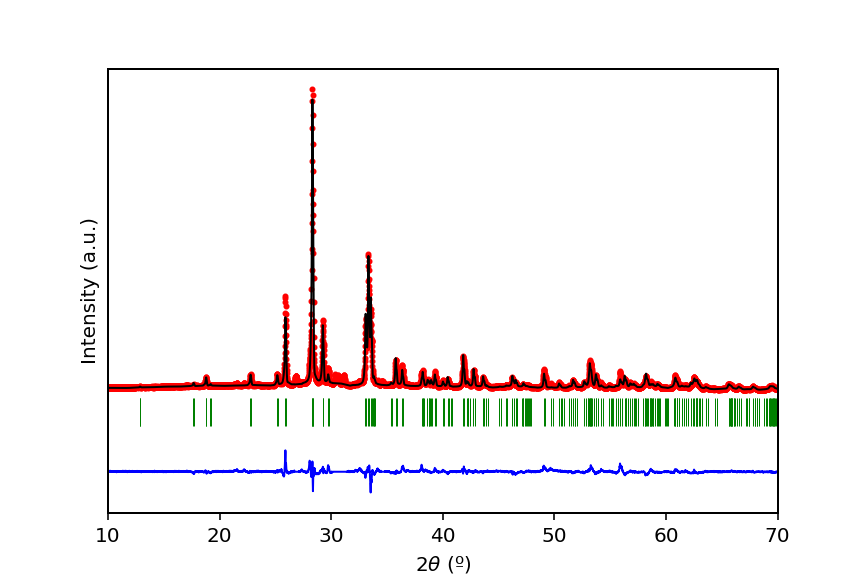} 
    \caption{\RaggedRight Experimental XRD pattern (red dots), calculated XRD pattern based on the structure obtained from Rietveld refinement (black line) and difference between the two curves (blue line) of the \ce{SrMn_{0.875}Cd_{0.125}Ge2O6} compound. Green tick marks indicate the position of the Bragg reflections for the $C2/c$ space group.}
    \label{fig:XRDSrMn0.875}
\end{figure*}

\begin{figure*}[h]
    \centering
    \includegraphics[width=1\textwidth,height=10cm]{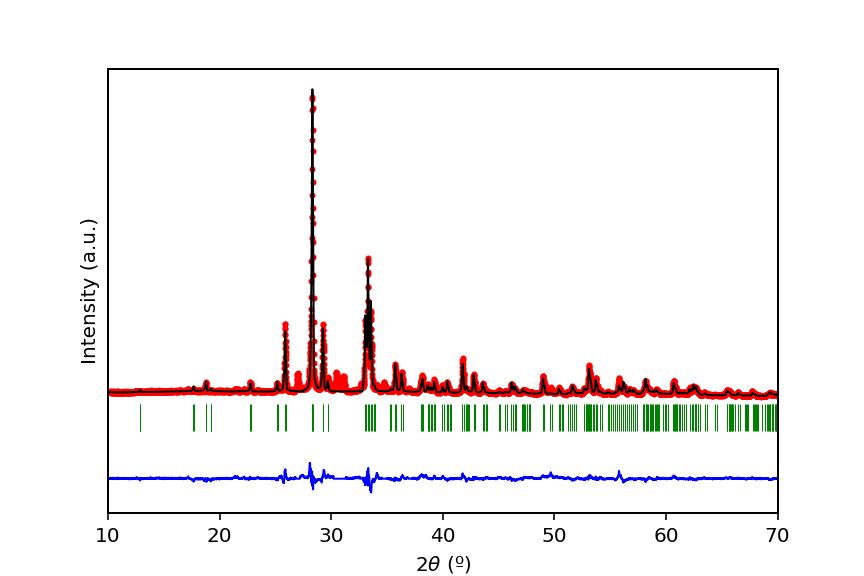} 
    \caption{\RaggedRight Experimental XRD pattern (red dots), calculated XRD pattern based on the structure obtained from Rietveld refinement (black line) and difference between the two curves (blue line) of the \ce{SrMn_{0.75}Cd_{0.25}Ge2O6} compound. Green tick marks indicate the position of the Bragg reflections for the $C2/c$ space group.}
    \label{fig:XRDSrMn0.75}
\end{figure*}

\vspace{2cm}

\begin{table}
\centering
\renewcommand{\arraystretch}{1.8} 
\setlength{\tabcolsep}{6pt} 
\caption{\RaggedRight Refined structural parameters for \ce{Ca_{0.875}Cd_{0.125}MnGe2O6} compound. Space group $C2/c$ (\#15); $\chi^{2}=4.42$; $R_{wp}=5.51$; $R_{Bragg}=7.21$.}
\label{table:PositionsCa0.875}
 \begin{tabular}{|c c c c c|}
\hline\hline
  & $x/a$  & $y/b$ & $z/c$ & $U_{iso} (\AA^{2})$   \\
\hline \hline
Ca/Cd  & 0.0 & 0.302(1) & 0.25 & 0.050(1) \\
\hline
Mn  & 0.0 & 0.906(1) & 0.25 & 0.032(1) \\
\hline
Ge  & 0.288(1) & 0.095(1) & 0.233(1) & 0.038(1) \\
\hline
O(1)  & 0.108(1) & 0.103(1) & 0.138(1) & 0.035(1) \\
\hline
O(2)  & 0.379(1) & 0.256(1) & 0.363(1) & 0.028(1) \\
\hline
O(3)  & 0.350(1) & 0.025(1) & 0.977(1) & 0.029(1) \\
\hline
\end{tabular}
\end{table}

\begin{table}
\centering
\renewcommand{\arraystretch}{1.8} 
\setlength{\tabcolsep}{6pt} 
\caption{\RaggedRight Refined structural parameters for \ce{Ca_{0.75}Cd_{0.25}MnGe2O6}. Space group $C2/c$ (\#15); $\chi^{2}=3.33$; $R_{wp}=6.19$; $R_{Bragg}=6.97$.}
\label{table:PositionsCa0.75}
 \begin{tabular}{|c c c c c|}
\hline\hline
  & $x/a$  & $y/b$ & $z/c$ & $U_{iso} (\AA^{2})$   \\
\hline
Ca/Cd  & 0.0 & 0.299(1) & 0.25 & 0.073(1) \\
\hline
Mn  & 0.0 & 0.910(1) & 0.25 & 0.033(1) \\
\hline
Ge  & 0.288(1) & 0.096(1) & 0.230(1) & 0.051(1) \\
\hline
O(1)  & 0.111(1) & 0.093(1) & 0.141(1) & 0.032(2) \\
\hline
O(2)  & 0.375(1) & 0.252(1) & 0.346(1) & 0.058(3) \\
\hline
O(3)  & 0.358(1) & 0.041(1) & 0.974(1) & 0.046(3) \\
\hline \hline
\end{tabular}
\end{table}

\begin{table}
\centering
\renewcommand{\arraystretch}{1.8} 
\setlength{\tabcolsep}{6pt} 
\caption{\RaggedRight Refined structural parameters for \ce{CaMn_{0.875}Cd_{0.125}Ge2O6}. Space group $C2/c$ (\#15); $\chi^{2}=3.22$; $R_{wp}=5.40$; $R_{Bragg}=6.73$.}
\label{table:PositionsCaMn0.875}
 \begin{tabular}{|c c c c c|}
\hline\hline
  & $x/a$  & $y/b$ & $z/c$ & $U_{iso} (\AA^{2})$   \\
\hline
Ca  & 0.0 & 0.304(1) & 0.25 & 0.036(1) \\
\hline
Mn/Cd  & 0.0 & 0.907(1) & 0.25 & 0.070(1) \\
\hline
Ge  & 0.287(1) & 0.097(1) & 0.233(1) & 0.047(1) \\
\hline
O(1)  & 0.115(1) & 0.104(1) & 0.148(1) & 0.053(2) \\
\hline
O(2)  & 0.368(1) & 0.253(1) & 0.358(1) & 0.023(1) \\
\hline
O(3)  & 0.358(1) & 0.023(1) & 0.983(1) & 0.049(2)  \\
\hline \hline
\end{tabular}
\end{table}

\begin{table}
\centering
\renewcommand{\arraystretch}{1.8} 
\setlength{\tabcolsep}{5.5pt} 
\caption{\RaggedRight Refined structural parameters for \ce{CaMn_{0.75}Cd_{0.25}Ge2O6}. Space group $C2/c$ (\#15); $\chi^{2}=7.73$; $R_{wp}=5.75$; $R_{Bragg}=12.3$.}
\label{table:PositionsCaMn0.75}
 \begin{tabular}{|c c c c c|}
\hline\hline
  & $x/a$  & $y/b$ & $z/c$ & $U_{iso} (\AA^{2})$   \\
\hline
Ca  & 0.0 & 0.293(1) & 0.25 & 0.0003(1) \\
\hline
Mn/Cd  & 0.0 & 0.902(1) & 0.25 & 0.005(1) \\
\hline
Ge  & 0.288(1) & 0.096(1) & 0.234(1) & 0.001(1) \\
\hline
O(1)  & 0.112(1) & 0.097(1) & 0.151(1) &  0.009(2) \\
\hline
O(2)  & 0.367(1) & 0.263(1) & 0.372(1) &  0.008(3) \\
\hline
O(3)  & 0.378(1) & 0.034(1) & 0.981(1) & 0.08(2)  \\
\hline \hline
\end{tabular}
\end{table}

\begin{table}
\centering
\renewcommand{\arraystretch}{1.8} 
\setlength{\tabcolsep}{6pt} 
\caption{\RaggedRight Refined structural parameters for the \ce{Sr_{0.875}Cd_{0.125}MnGe2O6} compound. Space group $C2/c$ (\#15); $\chi^{2}=7.21$; $R_{wp}=9.67$; $R_{Bragg}=15.1$.}
\label{table:PositionsSr0.875}
 \begin{tabular}{|c c c c c|}
\hline\hline
  & $x/a$  & $y/b$ & $z/c$ & $U_{iso} (\AA^{2})$   \\
\hline \hline
Sr/Cd  & 0.0 & 0.306(1) & 0.25 & 0.039(1) \\
\hline
Mn  & 0.0 & 0.907(1) & 0.25 & 0.033(1) \\
\hline
Ge  & 0.284(1) & 0.090(1) & 0.217(1)  & 0.053(1) \\
\hline
O(1)  & 0.108(1) & 0.079(1) & 0.123(1) & 0.030(4) \\
\hline
O(2)  & 0.391(1) & 0.241(1) & 0.367(2) & 0.099(6) \\
\hline
O(3)  & 0.338(1) & 0.013(1) & 0.990(2) & 0.036(5) \\
\hline \hline
\end{tabular}
\end{table}

\begin{table}
\centering
\renewcommand{\arraystretch}{1.8} 
\setlength{\tabcolsep}{6pt} 
\caption{\RaggedRight Refined structural parameters for the \ce{Sr_{0.75}Cd_{0.25}MnGe2O6} compound. Space group $C2/c$ (\#15); $\chi^{2}=4.45$; $R_{wp}=7.26$; $R_{Bragg}=10.7$.}
\label{table:PositionsSr0.75}
 \begin{tabular}{|c c c c c|}
 \hline \hline
  & $x/a$  & $y/b$ & $z/c$ & $U_{iso} (\AA^{2})$   \\
\hline \hline
Sr/Cd  & 0.0 &  0.303(1)  & 0.25 &  0.081(1) \\
\hline
Mn  & 0.0 & 0.908(1) & 0.25 & 0.052(1) \\
\hline
Ge  & 0.285(1) & 0.093(1) & 0.224(1) & 0.067(1) \\
\hline
O(1)  & 0.105(1) & 0.087(1) & 0.135(1) & 0.044(3) \\
\hline
O(2)  & 0.376(1) & 0.258(1) & 0.339(1) & 0.103(4) \\
\hline
O(3)  & 0.347(1) & 0.037(1) & 0.995(1) & 0.072(4)  \\
\hline \hline
\end{tabular}
\end{table}

\begin{table}
\centering
\renewcommand{\arraystretch}{1.75} 
\setlength{\tabcolsep}{4.8pt} 
\caption{\RaggedRight Refined structural parameters for \ce{SrMn_{0.875}Cd_{0.125}Ge2O6}compound. Space group $C2/c$ (\#15); $\chi^{2}=6.46$; $R_{wp}=8.46$; $R_{Bragg}=12.2$.}
\label{table:PositionsSrMn0.875}
 \begin{tabular}{|c c c c c|}
\hline\hline
  & $x/a$  & $y/b$ & $z/c$ & $U_{iso} (\AA^{2})$   \\
\hline \hline
Sr  & 0.0 & 0.3057(3) & 0.25 & 0.0426(11) \\
\hline
Mn/Cd  & 0.0 & 0.9065(4) & 0.25 & 0.0572(15) \\
\hline
Ge  & 0.2831(2) & 0.092(1) & 0.214(1) & 0.054(1) \\
\hline
O(1)  & 0.101(1) & 0.089(1) & 0.132(1) & 0.019(3) \\
\hline
O(2)  & 0.382(1) & 0.248(1) & 0.357(2) & 0.087(5) \\
\hline
O(3)  & 0.343(1) & 0.009(1) & 0.987(1) & 0.019(4)  \\
\hline \hline
\end{tabular}
\end{table}

\begin{table}[]
\centering
\renewcommand{\arraystretch}{1.8} 
\setlength{\tabcolsep}{6pt} 
\caption{\RaggedRight \RaggedRight Refined structural parameters for \ce{SrMn_{0.75}Cd_{0.25}Ge2O6} compound. Space group $C2/c$ (\#15); $\chi^{2}=5.87$; $R_{wp}=9.43$; $R_{Bragg}=12.9$.}
\label{table:PositionsSrMn0.75}
 \begin{tabular}{|c c c c c|}
\hline\hline
  & $x/a$  & $y/b$ & $z/c$ & $U_{iso} (\AA^{2})$   \\
\hline \hline
Sr  & 0.0 & 0.307(1) & 0.25 & 0.052(1) \\
\hline
Mn/Cd  & 0.0 & 0.906(1) & 0.25 & 0.087(1) \\
\hline
Ge  & 0.284(1) & 0.092(1) & 0.214(1) & 0.066(1) \\
\hline
O(1)  & 0.095(1) & 0.090(1) & 0.135(1) & 0.029(3) \\
\hline
O(2)  & 0.369(1) & 0.240(1) & 0.360(1) & 0.054(5) \\
\hline
O(3)  & 0.343(1) & 0.009(1) & 0.982(1) & 0.029(4)  \\
\hline
\end{tabular}
\end{table}

\begin{table}
\centering
\renewcommand{\arraystretch}{1.6} 
\caption{\RaggedRight Refined lattice parameters for the $C2/c$  \ce{\textit{A}MnGe2O6} (\textit{A} = Ca, Sr) clinopyroxene compounds where Cd substitutes $A$ or Mn ions.} 
\label{table:Parameters}
 \scalebox{0.9}{
 \begin{tabular}{|l c c c c|}
\hline\hline
  ~~~~~~~~Compound & $a$ (\AA) & $b$ (\AA) & $c$ (\AA) & $\beta$ ($^{\circ}$) \\
  \hline \hline 
\ce{Ca_{0.875}Cd_{0.125}MnGe2O6}    &  10.247(1) &  9.161(1) &  5.454(1) &   104.247(1)  \\
\hline 
\ce{Ca_{0.75}Cd_{0.25}MnGe2O6}        &  10.222(1) & 9.198(1) &  5.439(1) &   104.083(1) \\
\hline 
\ce{CaMn_{0.875}Cd_{0.125}Ge2O6}    &  10.272(1) & 9.181(1) &  5.466(1) &    104.167(1)  \\
\hline 
 \ce{CaMn_{0.75}Cd_{0.25}Ge2O6}   & 10.265(1) &  9.211(1) &  5.458(1) &   104.042(1)  \\
 \hline 
 \ce{Sr_{0.875}Cd_{0.125}MnGe2O6}    & 10.344(1) &   9.415(1) &  5.507(1) &  104.736(1) \\ 
\hline     
\ce{Sr_{0.75}Cd_{0.25}MnGe2O6}        & 10.329(1) &  9.423(1) &  5.500(1) & 104.762(1) \\
\hline 
\ce{SrMn_{0.875}Cd_{0.125}Ge2O6}    &  10.356(1) &  9.429(1) &  5.511(1) &  104.638(1)  \\ 
\hline
\ce{SrMn_{0.75}Cd_{0.25}Ge2O6}        &  10.378(1) &  9.455(1) &  5.516(1) &  104.543(1) \\
 \hline \hline 
\end{tabular}}
\end{table}

\FloatBarrier

\section{Electronic Band Dispersion of the Cd Doped Systems}
\label{Sec:A3}
\setcounter{table}{0}
\setcounter{figure}{0}

\begin{figure}
\centering
\includegraphics[width=1.1\linewidth]{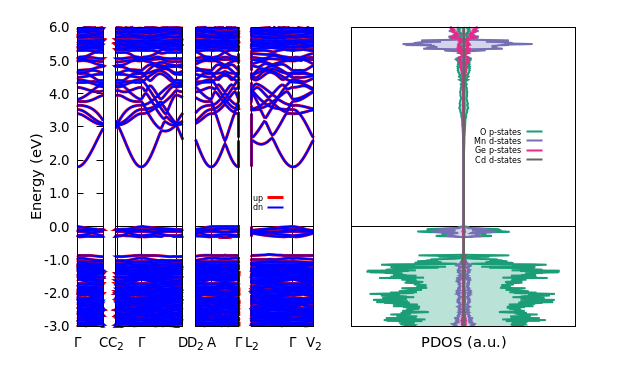}
\caption[\RaggedRight Electronic band structure (left) and density of states (right) for \ce{CaMn0.75Mn0.25Ge2O6} with antiferromagnetic ordering.]{Electronic bandstructure and density of states for \ce{CaMn_{0.75}Cd_{0.25}Ge2O6} 
with antiferromagnetic ordering.}
\label{fig:Ca2Mn2Ge4O12AFM_bands}
\end{figure}

\begin{widetext}
Figures \ref{fig:FigSrBands_DOS} to \ref{fig:Ca2Mn2Ge4O12AFM_bands} present the band structures  and the density of states for the \ce{\textit{A}MnGe2O6} (\textit{A} = Ca, Sr) clinopyroxene compounds where $A$ or Mn  atoms are substituted by Cd ions.

\begin{figure*}[]
    \centering
    \subfloat{\includegraphics[width=0.48\textwidth]{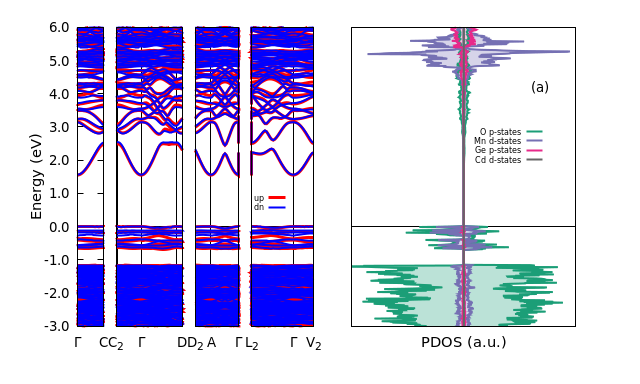}
    }\hfill
    \subfloat
        {\includegraphics[width=0.48\textwidth]{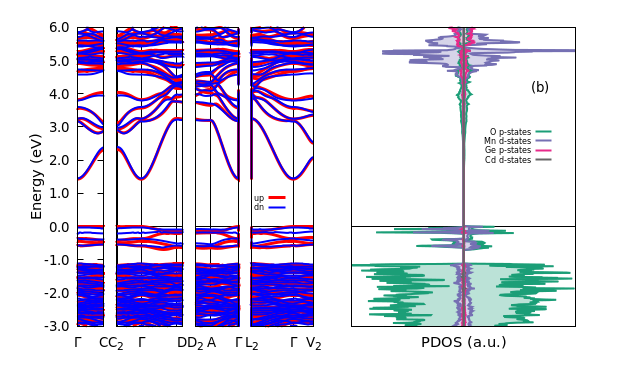}
    } \\
    \subfloat
        {\includegraphics[width=0.48\textwidth]{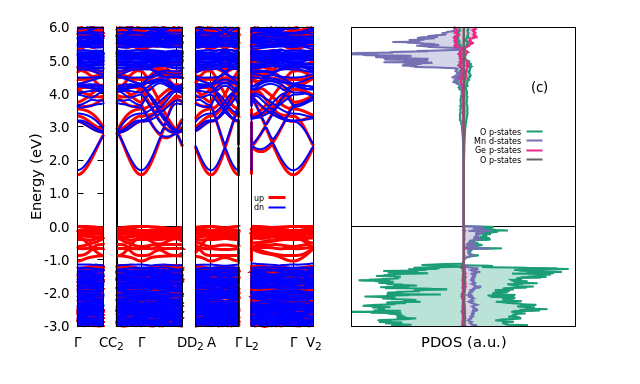}
    }\hfill
    \subfloat
        {\includegraphics[width=0.48\textwidth]{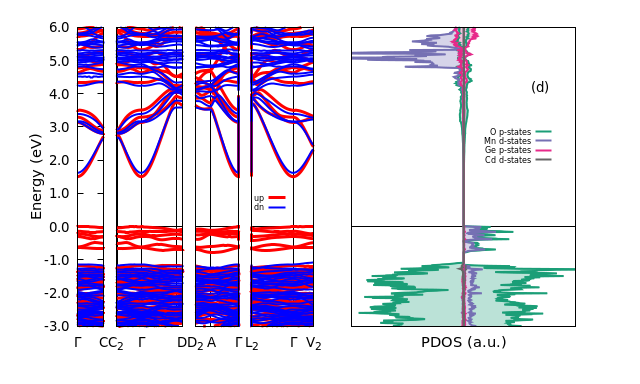}
    } 
    \caption{\RaggedRight Electronic band structure (left) and density of states (right) for: (a) \ce{Sr_{0.875}Cd_{0.125}MnGe2O6} with antiferromagnetic ordering; (b) \ce{Sr_{0.75}Cd_{0.25}MnGe2O6} with antiferromagnetic ordering; (c) \ce{SrMn_{0.875}Cd_{0.125}Ge2O6} with ferromagnetic ordering; (d) \ce{SrMn_{0.75}Cd_{0.25}Ge2O6} with ferromagnetic ordering.}
    \label{fig:FigSrBands_DOS}
\end{figure*}

\begin{figure*}[]
\centering
    \label{fig:FigCaBands_DOS}
    \subfloat
        {\includegraphics[width=0.48\textwidth]{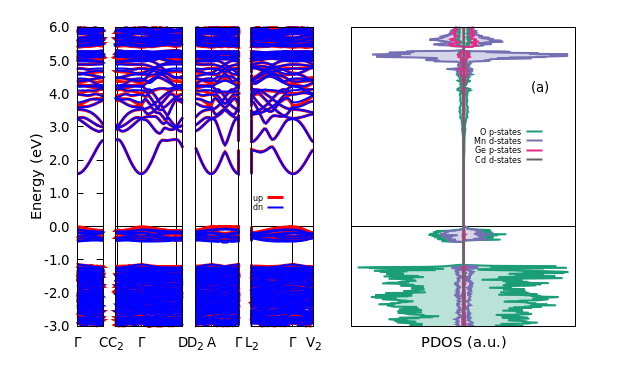}
    }\hfill
    \subfloat
        {\includegraphics[width=0.48\textwidth]{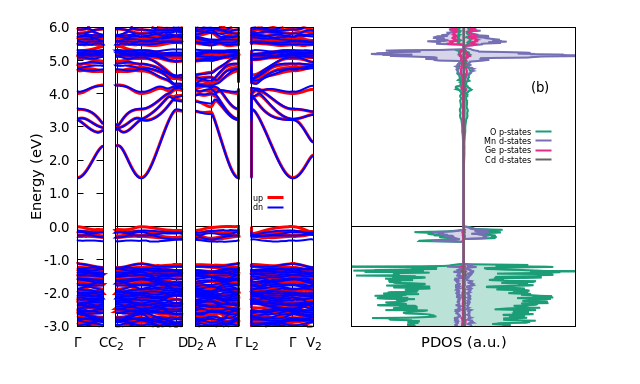}
    } \\
    \subfloat
        {\includegraphics[width=0.48\textwidth]{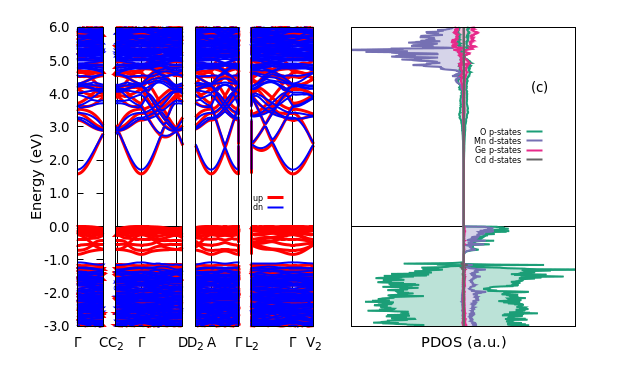}
    }\hfill
    \subfloat
        {\includegraphics[width=0.48\textwidth]{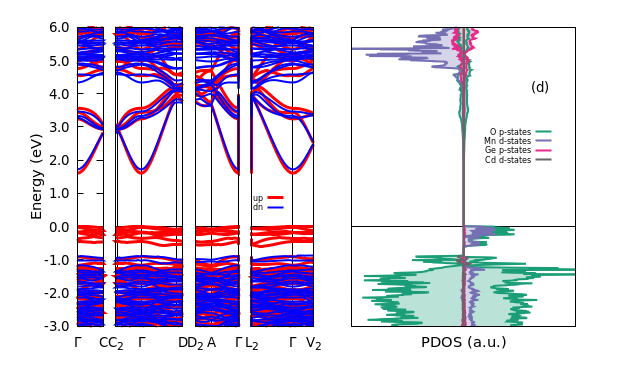}
    } 
    \caption{\RaggedRight Electronic band structure (left) and density of states (right) for: (a) \ce{Ca_{0.875}Cd_{0.125}MnGe2O6} with antiferromagnetic ordering; (b) \ce{Ca_{0.75}Cd_{0.25}MnGe2O6} with antiferromagnetic ordering; (c) \ce{CaMn_{0.875}Cd_{0.125}Ge2O6} with ferromagnetic ordering; (d) \ce{CaMn_{0.75}Cd_{0.25}Ge2O6} with ferromagnetic ordering.}
\end{figure*}
\end{widetext}

\vspace{2cm}

\section{Electric Field Gradient Parameters}
\label{Sec:A2}
\setcounter{table}{0}
\setcounter{figure}{0}

The calculated electric field gradient (EFG) parameters are presented in table \ref{table:EFG}. From these results we can observe that the EFGs at the Ge nuclei  were only slightly affected by the Cd substitution, as evidenced by the fact that the symmetry of the Ge sites remained quite similar to \ce{SrMnGe2O6} and \ce{CaMnGe2O6} basic compounds after Cd substitution. The same cannot be said for the Sr/Ca and Mn sites, which in many cases deviated substantially.

\begin{widetext}
\begin{table*}

	\centering
    \renewcommand{\arraystretch}{1.6} 
    \setlength{\tabcolsep}{3pt} 
	\caption{\RaggedRight Computational EFG parameters at the Sr, Ca, Mn, Ge, and Cd nuclei  for  $\rm{SrMnGe_2O_6}$,  $\rm {SrMnGe_2O_6}$:$\rm{Cd_{Sr}}$,  and $\rm {SrMnGe_2O_6}$:$\rm{Cd_{Mn}}$, $\rm{CaMnGe_2O_6}$,  $\rm{CaMnGe_2O_6}$:$\rm{Cd_{Ca}}$,  and 
	$\rm{SrMnGe_2O_6}$:$\rm{Cd_{Mn}}$ compounds. $|V_{zz}|$ is given in V/\AA$^{2}$ units.}
	\begin{center}
 \scalebox{0.95}{	
		\begin{tabular}{lccccccccccccccc}
\hline	\hline
	&&&&&&&&&& & & & &\\	[-5mm]
System & EFG & Cd & Sr(1) & Sr(2) & Sr(3) & Sr(4) & Mn(1) & Mn(2) & Mn(3) & Mn(4) & Ge(1) & Ge(2) & Ge(3) & Ge(4) \\
	&&&&&&&&&& & & & &\\	[-5mm]
\hline	\hline
\multicolumn{1}{l}{\multirow{2}{*}{$\rm SrMnGe_2O_6$}} & \multicolumn{1}{c}{$|V_{zz}|$} & - & 64 & - & - & - & 14 & - & - &   -  & 99 & - & - & - \\ 
\multicolumn{1}{l}{} & \multicolumn{1}{c}{$\eta$} & - & 0.54 & - & - & - & 0.32 & - & - &   - & 0.70 & - & - & -  \\  
 \hline
\multicolumn{1}{l}{\multirow{2}{*}{$\rm Sr_{0.875}Cd_{0.125}MnGe_2O_6$}} & \multicolumn{1}{c}{$|V_{zz}|$} & 108 & 61 & 62 & 63 & 66 & 10 & 19 & 17 & 12 & 101 & 97 & 102 & 101 \\ 
\multicolumn{1}{l}{} & \multicolumn{1}{c}{$\eta$} & 0.14 & 0.46 & 0.71 & 0.53 & 0.83 & 0.64 & 0.98 & 0.44 &  0.84 & 0.60 & 0.64 & 0.64 & 0.69 \\  
\hline
\multicolumn{1}{l}{\multirow{2}{*}{$\rm Sr_{0.75}Cd_{0.25}MnGe_2O_6$}} & \multicolumn{1}{c}{$|V_{zz}|$} & 103 & 59 & 65 & 64 & - & 19 & 17 & 14 &   15  & 99 & 103 & 91 & 85\\ 
\multicolumn{1}{l}{} & \multicolumn{1}{c}{$\eta$} & 0.14 & 0.88 & 0.82 & 0.66 & - & 0.42 & 0.75&0.60 &  0.16 & 0.55 & 0.64 & 0.55 &  0.67\\  
 \hline
\multicolumn{1}{l}{\multirow{2}{*}{$\rm {SrMn_{0.875}Cd_{0.125}Ge_2O_6}$}} & \multicolumn{1}{c}{$|V_{zz}|$} & 21 & 57 & 67 & 64 & 68 & 14 & 14 & 11 & 14 & 99 & 99 & 101 & 100 \\ 
\multicolumn{1}{l}{} & \multicolumn{1}{c}{$\eta$} & 0.74 & 0.47 & 0.63 & 0.57 & 0.59 & 0.28 & 0.46&0.63 & 0.35 & 0.71 & 0.69 & 0.69 & 0.69\\  
\hline 
\multicolumn{1}{l}{\multirow{2}{*}{$\rm {SrMn_{0.75}Cd_{0.25}Ge_2O_6}$}} & \multicolumn{1}{c}{$|V_{zz}|$} & 22 & 59 & 64 & 69 & 66 & 15 & 10 & 15 &   -  & 99 & 101 & 100 & 102\\ 
\multicolumn{1}{l}{} & \multicolumn{1}{c}{$\eta$} & 0.67 & 0.56 & 0.59 & 0.63 & 0.64& 0.58 & 0.61 & 0.23 &  -  & 0.71 & 0.67 & 0.73 & 0.69\\  
\hline \hline
	&&&&&&&&&& & & & &\\	[-4mm]
System & EFG & Cd & Ca(1) & Ca(2) & Ca(3) & Ca(4) & Mn(1) & Mn(2) & Mn(3) & Mn(4) & Ge(1) & Ge(2) & Ge(3) & Ge(4) \\
	&&&&&&&&&& & & & &\\	[-5mm]
\hline	\hline
\multicolumn{1}{l}{\multirow{2}{*}{$\rm CaMnGe_2O_6$}} & \multicolumn{1}{c}{$|V_{zz}|$} & - & 34 & - & - & - & 8 & - & - &   -  & 96 & - & - & - \\ 
\multicolumn{1}{l}{} & \multicolumn{1}{c}{$\eta$} & - & 0.52 & - & - & - & 0.49 & - & - &   -  & 0.64 & - & - & -\\  
 \hline
\multicolumn{1}{l}{\multirow{2}{*}{$\rm Ca_{0.875}Cd_{0.125}MnGe_2O_6$}} & \multicolumn{1}{c}{$|V_{zz}|$} & 109 & 33 & 35 & 34 & 38 & 5 & 8 & 11 &   9 & 99 & 97 & 98 & 99 \\ 
\multicolumn{1}{l}{} & \multicolumn{1}{c}{$\eta$} & 0.01 & 0.45 & 0.34 & 0.45 & 0.33 & 0.43 & 0.30 &0.69 &  0.93 & 0.52 & 0.60 & 0.60 & 0.60 \\  
\hline
\multicolumn{1}{l}{\multirow{2}{*}{$\rm Ca_{0.75}Cd_{0.25}MnGe_2O_6$}} & \multicolumn{1}{c}{$|V_{zz}|$} & 108 & 35 &38 & 35 & - & 2 & 12 & 11 &   8&  99 & 100 & 96 &  86\\ 
\multicolumn{1}{l}{} & \multicolumn{1}{c}{$\eta$} & 0.02 & 0.13 & 0.27 & 0.29 & - & 0.09 & 0.11&0.15 &  0.95  & 0.50 & 0.58 & 0.44 & 0.67\\  
\hline
\multicolumn{1}{l}{\multirow{2}{*}{$\rm {CaMn_{0.875}Cd_{0.125}Ge_2O_6}$}} & \multicolumn{1}{c}{$|V_{zz}|$} & 12 & 30 & 37 & 36 & 36 & 7 & 7 & 10 & 7 & 98 & 97 & 99 & 97\\ 
\multicolumn{1}{l}{} & \multicolumn{1}{c}{$\eta$} & 0.21 & 0.42 & 0.41 & 0.38 & 0.39 & 0.76 & 0.92 & 0.34 & 0.82 & 0.63 & 0.61 & 0.60 & 0.66\\  
\hline 
\multicolumn{1}{l}{\multirow{2}{*}{$\rm {CaMn_{0.75}Cd_{0.25}Ge_2O_6}$}} & \multicolumn{1}{c}{$|V_{zz}|$} & 12 & 31 & 37 & 38 & 37 & 7 & 10 & 8 & - & 97 & 99 & 98 & 100\\ 
\multicolumn{1}{l}{} & \multicolumn{1}{c}{$\eta$} & 0.31  & 0.42 & 0.39 & 0.40 & 0.39 & 0.37 & 0.38 & 0.96 & - & 0.63 & 0.58 & 0.64 & 0.59 \\  
 \hline \hline
\end{tabular}}
\label{table:EFG}
	\end{center}
\end{table*} 

\end{widetext}


\begin{thebibliography}{51}
\expandafter\ifx\csname natexlab\endcsname\relax\def\natexlab#1{#1}\fi
\expandafter\ifx\csname bibnamefont\endcsname\relax
  \def\bibnamefont#1{#1}\fi
\expandafter\ifx\csname bibfnamefont\endcsname\relax
  \def\bibfnamefont#1{#1}\fi
\expandafter\ifx\csname citenamefont\endcsname\relax
  \def\citenamefont#1{#1}\fi
\expandafter\ifx\csname url\endcsname\relax
  \def\url#1{\texttt{#1}}\fi
\expandafter\ifx\csname urlprefix\endcsname\relax\def\urlprefix{URL }\fi
\providecommand{\bibinfo}[2]{#2}
\providecommand{\eprint}[2][]{\url{#2}}



\bibitem[{\citenamefont{Streltsov and Khomskii}(2008)}]{Streltsov2008}
\bibinfo{author}{\bibfnamefont{S.~V.} \bibnamefont{Streltsov}} \bibnamefont{and} \bibinfo{author}{\bibfnamefont{D.~I.} \bibnamefont{Khomskii}}, \bibinfo{journal}{Phys. Rev. B} \textbf{\bibinfo{volume}{77}}, \bibinfo{pages}{064405} (\bibinfo{year}{2008}).

\bibitem[{\citenamefont{Ding et~al.}(2016{\natexlab{a}})\citenamefont{Ding, Colin, Darie, and Bordet}}]{Ding_JMCC2016}
\bibinfo{author}{\bibfnamefont{L.}~\bibnamefont{Ding}}, \bibinfo{author}{\bibfnamefont{C.~V.} \bibnamefont{Colin}}, \bibinfo{author}{\bibfnamefont{C.}~\bibnamefont{Darie}}, \bibnamefont{and} \bibinfo{author}{\bibfnamefont{P.}~\bibnamefont{Bordet}}, \bibinfo{journal}{J. Mater. Chem. C} \textbf{\bibinfo{volume}{4}}, \bibinfo{pages}{4236} (\bibinfo{year}{2016}{\natexlab{a}}).

\bibitem[{\citenamefont{Colin et~al.}(2020)\citenamefont{Colin, Ding, Ressouche, Robert, Terada, Gay, Lejay, Simonet, Darie, Bordet et~al.}}]{PhysRevB.101.235109}
\bibinfo{author}{\bibfnamefont{C.~V.} \bibnamefont{Colin}}, \bibinfo{author}{\bibfnamefont{L.}~\bibnamefont{Ding}}, \bibinfo{author}{\bibfnamefont{E.}~\bibnamefont{Ressouche}}, \bibinfo{author}{\bibfnamefont{J.}~\bibnamefont{Robert}}, \bibinfo{author}{\bibfnamefont{N.}~\bibnamefont{Terada}}, \bibinfo{author}{\bibfnamefont{F.}~\bibnamefont{Gay}}, \bibinfo{author}{\bibfnamefont{P.}~\bibnamefont{Lejay}}, \bibinfo{author}{\bibfnamefont{V.}~\bibnamefont{Simonet}}, \bibinfo{author}{\bibfnamefont{C.}~\bibnamefont{Darie}}, \bibinfo{author}{\bibfnamefont{P.}~\bibnamefont{Bordet}}, \bibnamefont{et~al.}, \bibinfo{journal}{Phys. Rev. B} \textbf{\bibinfo{volume}{101}}, \bibinfo{pages}{235109} (\bibinfo{year}{2020}).

\bibitem[{\citenamefont{Jodlauk et~al.}(2007)\citenamefont{Jodlauk, Becker, Mydosh, Khomskii, Lorenz, Streltsov, Hezel, and Bohat{\'{y}}}}]{Jodlauk2007}
\bibinfo{author}{\bibfnamefont{S.}~\bibnamefont{Jodlauk}}, \bibinfo{author}{\bibfnamefont{P.}~\bibnamefont{Becker}}, \bibinfo{author}{\bibfnamefont{J.~A.} \bibnamefont{Mydosh}}, \bibinfo{author}{\bibfnamefont{D.~I.} \bibnamefont{Khomskii}}, \bibinfo{author}{\bibfnamefont{T.}~\bibnamefont{Lorenz}}, \bibinfo{author}{\bibfnamefont{S.~V.} \bibnamefont{Streltsov}}, \bibinfo{author}{\bibfnamefont{D.~C.} \bibnamefont{Hezel}}, \bibnamefont{and} \bibinfo{author}{\bibfnamefont{L.}~\bibnamefont{Bohat{\'{y}}}}, \bibinfo{journal}{J. Phys. Condens. Matter.} \textbf{\bibinfo{volume}{19}}, \bibinfo{pages}{432201} (\bibinfo{year}{2007}).

\bibitem[{\citenamefont{Kim et~al.}(2012)\citenamefont{Kim, Jeon, Patil, Patil, N{\'{e}}nert, and Kim}}]{Kim2012}
\bibinfo{author}{\bibfnamefont{I.}~\bibnamefont{Kim}}, \bibinfo{author}{\bibfnamefont{B.-G.} \bibnamefont{Jeon}}, \bibinfo{author}{\bibfnamefont{D.}~\bibnamefont{Patil}}, \bibinfo{author}{\bibfnamefont{S.}~\bibnamefont{Patil}}, \bibinfo{author}{\bibfnamefont{G.}~\bibnamefont{N{\'{e}}nert}}, \bibnamefont{and} \bibinfo{author}{\bibfnamefont{K.~H.} \bibnamefont{Kim}}, \bibinfo{journal}{J. Phys. Condens. Matter.} \textbf{\bibinfo{volume}{24}}, \bibinfo{pages}{306001} (\bibinfo{year}{2012}).

\bibitem[{\citenamefont{Ding et~al.}(2016{\natexlab{b}})\citenamefont{Ding, Colin, Darie, Robert, Gay, and Bordet}}]{Ding_PRB2016}
\bibinfo{author}{\bibfnamefont{L.}~\bibnamefont{Ding}}, \bibinfo{author}{\bibfnamefont{C.~V.} \bibnamefont{Colin}}, \bibinfo{author}{\bibfnamefont{C.}~\bibnamefont{Darie}}, \bibinfo{author}{\bibfnamefont{J.}~\bibnamefont{Robert}}, \bibinfo{author}{\bibfnamefont{F.}~\bibnamefont{Gay}}, \bibnamefont{and} \bibinfo{author}{\bibfnamefont{P.}~\bibnamefont{Bordet}}, \bibinfo{journal}{Phys. Rev. B} \textbf{\bibinfo{volume}{93}}, \bibinfo{pages}{064423} (\bibinfo{year}{2016}{\natexlab{b}}).

\bibitem[{\citenamefont{Legesse et~al.}(2018)\citenamefont{Legesse, Park, El~Mellouhi, Rashkeev, Kais, and Alharbi}}]{https://doi.org/10.1002/cphc.201701155}
\bibinfo{author}{\bibfnamefont{M.}~\bibnamefont{Legesse}}, \bibinfo{author}{\bibfnamefont{H.}~\bibnamefont{Park}}, \bibinfo{author}{\bibfnamefont{F.}~\bibnamefont{El~Mellouhi}}, \bibinfo{author}{\bibfnamefont{S.~N.} \bibnamefont{Rashkeev}}, \bibinfo{author}{\bibfnamefont{S.}~\bibnamefont{Kais}}, \bibnamefont{and} \bibinfo{author}{\bibfnamefont{F.~H.} \bibnamefont{Alharbi}}, \bibinfo{journal}{Chem. Phys. Chem.} \textbf{\bibinfo{volume}{19}}, \bibinfo{pages}{943} (\bibinfo{year}{2018}).

\bibitem[{\citenamefont{Torres et~al.}(2019)\citenamefont{Torres, Luque, Tortajada, and Arroyo-de Dompablo}}]{Torres2019}
\bibinfo{author}{\bibfnamefont{A.}~\bibnamefont{Torres}}, \bibinfo{author}{\bibfnamefont{F.~J.} \bibnamefont{Luque}}, \bibinfo{author}{\bibfnamefont{J.}~\bibnamefont{Tortajada}}, \bibnamefont{and} \bibinfo{author}{\bibfnamefont{M.~E.} \bibnamefont{Arroyo-de Dompablo}}, \bibinfo{journal}{Sci. Rep.} \textbf{\bibinfo{volume}{9}}, \bibinfo{pages}{9644} (\bibinfo{year}{2019}).

\bibitem[{\citenamefont{Zhou et~al.}(2014)\citenamefont{Zhou, King, Scanlon, Sougrati, and Melot}}]{Zhou_2014}
\bibinfo{author}{\bibfnamefont{S.}~\bibnamefont{Zhou}}, \bibinfo{author}{\bibfnamefont{G.}~\bibnamefont{King}}, \bibinfo{author}{\bibfnamefont{D.~O.} \bibnamefont{Scanlon}}, \bibinfo{author}{\bibfnamefont{M.~T.} \bibnamefont{Sougrati}}, \bibnamefont{and} \bibinfo{author}{\bibfnamefont{B.~C.} \bibnamefont{Melot}}, \bibinfo{journal}{J. Electrochem. Soc.} \textbf{\bibinfo{volume}{161}}, \bibinfo{pages}{A1642} (\bibinfo{year}{2014}).

\bibitem[{\citenamefont{Jin et~al.}(2024)\citenamefont{Jin, Peng, Rutherford, Xu, Ni, Yang, Byeon, Xie, Zhou, Dai et~al.}}]{CoPyroJin}
\bibinfo{author}{\bibfnamefont{L.}~\bibnamefont{Jin}}, \bibinfo{author}{\bibfnamefont{S.}~\bibnamefont{Peng}}, \bibinfo{author}{\bibfnamefont{A.~N.} \bibnamefont{Rutherford}}, \bibinfo{author}{\bibfnamefont{X.}~\bibnamefont{Xu}}, \bibinfo{author}{\bibfnamefont{D.}~\bibnamefont{Ni}}, \bibinfo{author}{\bibfnamefont{C.}~\bibnamefont{Yang}}, \bibinfo{author}{\bibfnamefont{Y.~J.} \bibnamefont{Byeon}}, \bibinfo{author}{\bibfnamefont{W.}~\bibnamefont{Xie}}, \bibinfo{author}{\bibfnamefont{H.}~\bibnamefont{Zhou}}, \bibinfo{author}{\bibfnamefont{X.}~\bibnamefont{Dai}}, \bibnamefont{et~al.}, \bibinfo{journal}{Science Advances} \textbf{\bibinfo{volume}{10}}, \bibinfo{pages}{eadp4685} (\bibinfo{year}{2024}), \eprint{https://www.science.org/doi/pdf/10.1126/sciadv.adp4685}, \urlprefix\url{https://www.science.org/doi/abs/10.1126/sciadv.adp4685}.

\bibitem[{\citenamefont{Maksimov et~al.}(2024)\citenamefont{Maksimov, Ushakov, Gubkin, Redhammer, Winter, Kolesnikov, dos Santos, Gai, McGuire, Podlesnyak et~al.}}]{CoPyroPavel}
\bibinfo{author}{\bibfnamefont{P.~A.} \bibnamefont{Maksimov}}, \bibinfo{author}{\bibfnamefont{A.~V.} \bibnamefont{Ushakov}}, \bibinfo{author}{\bibfnamefont{A.~F.} \bibnamefont{Gubkin}}, \bibinfo{author}{\bibfnamefont{G.~J.} \bibnamefont{Redhammer}}, \bibinfo{author}{\bibfnamefont{S.~M.} \bibnamefont{Winter}}, \bibinfo{author}{\bibfnamefont{A.~I.} \bibnamefont{Kolesnikov}}, \bibinfo{author}{\bibfnamefont{A.~M.} \bibnamefont{dos Santos}}, \bibinfo{author}{\bibfnamefont{Z.}~\bibnamefont{Gai}}, \bibinfo{author}{\bibfnamefont{M.~A.} \bibnamefont{McGuire}}, \bibinfo{author}{\bibfnamefont{A.}~\bibnamefont{Podlesnyak}}, \bibnamefont{et~al.}, \bibinfo{journal}{Proceedings of the National Academy of Sciences} \textbf{\bibinfo{volume}{121}}, \bibinfo{pages}{e2409154121} (\bibinfo{year}{2024}), \eprint{https://www.pnas.org/doi/pdf/10.1073/pnas.2409154121}, \urlprefix\url{https://www.pnas.org/doi/abs/10.1073/pnas.2409154121}.

\bibitem[{\citenamefont{Temnikov et~al.}(2019)\citenamefont{Temnikov, Komleva, Pchelkina, and Streltsov}}]{Temnikov2019}
\bibinfo{author}{\bibfnamefont{F.~V.} \bibnamefont{Temnikov}}, \bibinfo{author}{\bibfnamefont{E.~V.} \bibnamefont{Komleva}}, \bibinfo{author}{\bibfnamefont{Z.~V.} \bibnamefont{Pchelkina}}, \bibnamefont{and} \bibinfo{author}{\bibfnamefont{S.~V.} \bibnamefont{Streltsov}}, \bibinfo{journal}{JETP Lett.} \textbf{\bibinfo{volume}{110}}, \bibinfo{pages}{595} (\bibinfo{year}{2019}).

\bibitem[{\citenamefont{Akter et~al.}(2025)\citenamefont{Akter, Islam, Hossain, and Rabu}}]{CMGOAkter}
\bibinfo{author}{\bibfnamefont{T.}~\bibnamefont{Akter}}, \bibinfo{author}{\bibfnamefont{J.}~\bibnamefont{Islam}}, \bibinfo{author}{\bibfnamefont{K.}~\bibnamefont{Hossain}}, \bibnamefont{and} \bibinfo{author}{\bibfnamefont{R.~A.} \bibnamefont{Rabu}}, \bibinfo{journal}{Heliyon} \textbf{\bibinfo{volume}{11}}, \bibinfo{pages}{e41315} (\bibinfo{year}{2025}), ISSN \bibinfo{issn}{2405-8440}, \urlprefix\url{https://www.sciencedirect.com/science/article/pii/S240584402417346X}.

\bibitem[{\citenamefont{Fakhera et~al.}(2023)\citenamefont{Fakhera, Hossain, Khanom, Hossain, and Ahmed}}]{SMGOFakhera}
\bibinfo{author}{\bibfnamefont{F.}~\bibnamefont{Fakhera}}, \bibinfo{author}{\bibfnamefont{K.}~\bibnamefont{Hossain}}, \bibinfo{author}{\bibfnamefont{M.~S.} \bibnamefont{Khanom}}, \bibinfo{author}{\bibfnamefont{M.~K.} \bibnamefont{Hossain}}, \bibnamefont{and} \bibinfo{author}{\bibfnamefont{F.}~\bibnamefont{Ahmed}}, \bibinfo{journal}{Journal of Physics and Chemistry of Solids} \textbf{\bibinfo{volume}{173}}, \bibinfo{pages}{111112} (\bibinfo{year}{2023}), ISSN \bibinfo{issn}{0022-3697}, \urlprefix\url{https://www.sciencedirect.com/science/article/pii/S0022369722005297}.

\bibitem[{\citenamefont{Giannozzi et~al.}(2009)\citenamefont{Giannozzi, Baroni, Bonini, Calandra, Car, Cavazzoni, Ceresoli, Chiarotti, Cococcioni, Dabo et~al.}}]{QE-2009}
\bibinfo{author}{\bibfnamefont{P.}~\bibnamefont{Giannozzi}}, \bibinfo{author}{\bibfnamefont{S.}~\bibnamefont{Baroni}}, \bibinfo{author}{\bibfnamefont{N.}~\bibnamefont{Bonini}}, \bibinfo{author}{\bibfnamefont{M.}~\bibnamefont{Calandra}}, \bibinfo{author}{\bibfnamefont{R.}~\bibnamefont{Car}}, \bibinfo{author}{\bibfnamefont{C.}~\bibnamefont{Cavazzoni}}, \bibinfo{author}{\bibfnamefont{D.}~\bibnamefont{Ceresoli}}, \bibinfo{author}{\bibfnamefont{G.~L.} \bibnamefont{Chiarotti}}, \bibinfo{author}{\bibfnamefont{M.}~\bibnamefont{Cococcioni}}, \bibinfo{author}{\bibfnamefont{I.}~\bibnamefont{Dabo}}, \bibnamefont{et~al.}, \bibinfo{journal}{J. Phys. Condens. Matter} \textbf{\bibinfo{volume}{21}}, \bibinfo{pages}{395502} (\bibinfo{year}{2009}).

\bibitem[{\citenamefont{Giannozzi et~al.}(2017)\citenamefont{Giannozzi, Andreussi, Brumme, Bunau, Nardelli, Calandra, Car, Cavazzoni, Ceresoli, Cococcioni et~al.}}]{QE-2017}
\bibinfo{author}{\bibfnamefont{P.}~\bibnamefont{Giannozzi}}, \bibinfo{author}{\bibfnamefont{O.}~\bibnamefont{Andreussi}}, \bibinfo{author}{\bibfnamefont{T.}~\bibnamefont{Brumme}}, \bibinfo{author}{\bibfnamefont{O.}~\bibnamefont{Bunau}}, \bibinfo{author}{\bibfnamefont{M.~B.} \bibnamefont{Nardelli}}, \bibinfo{author}{\bibfnamefont{M.}~\bibnamefont{Calandra}}, \bibinfo{author}{\bibfnamefont{R.}~\bibnamefont{Car}}, \bibinfo{author}{\bibfnamefont{C.}~\bibnamefont{Cavazzoni}}, \bibinfo{author}{\bibfnamefont{D.}~\bibnamefont{Ceresoli}}, \bibinfo{author}{\bibfnamefont{M.}~\bibnamefont{Cococcioni}}, \bibnamefont{et~al.}, \bibinfo{journal}{J. Phys. Condens. Matter} \textbf{\bibinfo{volume}{29}}, \bibinfo{pages}{465901} (\bibinfo{year}{2017}).

\bibitem[{\citenamefont{Bl\"ochl}(1994)}]{PhysRevB.50.17953}
\bibinfo{author}{\bibfnamefont{P.~E.} \bibnamefont{Bl\"ochl}}, \bibinfo{journal}{Phys. Rev. B} \textbf{\bibinfo{volume}{50}}, \bibinfo{pages}{17953} (\bibinfo{year}{1994}).

\bibitem[{\citenamefont{{Dal Corso}}(2014)}]{DALCORSO2014337}
\bibinfo{author}{\bibfnamefont{A.}~\bibnamefont{{Dal Corso}}}, \bibinfo{journal}{Comput. Mater. Sci.} \textbf{\bibinfo{volume}{95}}, \bibinfo{pages}{337} (\bibinfo{year}{2014}).

\bibitem[{\citenamefont{Perdew et~al.}(1996)\citenamefont{Perdew, Burke, and Ernzerhof}}]{Perdew1996}
\bibinfo{author}{\bibfnamefont{J.~P.} \bibnamefont{Perdew}}, \bibinfo{author}{\bibfnamefont{K.}~\bibnamefont{Burke}}, \bibnamefont{and} \bibinfo{author}{\bibfnamefont{M.}~\bibnamefont{Ernzerhof}}, \bibinfo{journal}{Phys. Rev. Lett.} \textbf{\bibinfo{volume}{77}}, \bibinfo{pages}{3865} (\bibinfo{year}{1996}).

\bibitem[{MP()}]{MP}
\bibinfo{note}{Data retrieved from the Materials Project for SrMn(GeO$_3$)$_2$ (mp-1208680) from database version v2022.10.28.}, \urlprefix\url{https://materialsproject.org/materials/mp-1208680?formula=SrMnGe2O6}.

\bibitem[{\citenamefont{Jain et~al.}(2013)\citenamefont{Jain, Ong, Hautier, Chen, Richards, Dacek, Cholia, Gunter, Skinner, Ceder et~al.}}]{doi:10.1063/1.4812323}
\bibinfo{author}{\bibfnamefont{A.}~\bibnamefont{Jain}}, \bibinfo{author}{\bibfnamefont{S.~P.} \bibnamefont{Ong}}, \bibinfo{author}{\bibfnamefont{G.}~\bibnamefont{Hautier}}, \bibinfo{author}{\bibfnamefont{W.}~\bibnamefont{Chen}}, \bibinfo{author}{\bibfnamefont{W.~D.} \bibnamefont{Richards}}, \bibinfo{author}{\bibfnamefont{S.}~\bibnamefont{Dacek}}, \bibinfo{author}{\bibfnamefont{S.}~\bibnamefont{Cholia}}, \bibinfo{author}{\bibfnamefont{D.}~\bibnamefont{Gunter}}, \bibinfo{author}{\bibfnamefont{D.}~\bibnamefont{Skinner}}, \bibinfo{author}{\bibfnamefont{G.}~\bibnamefont{Ceder}}, \bibnamefont{et~al.}, \bibinfo{journal}{APL Mater.} \textbf{\bibinfo{volume}{1}}, \bibinfo{pages}{011002} (\bibinfo{year}{2013}).

\bibitem[{\citenamefont{Cococcioni and de~Gironcoli}(2005)}]{PhysRevB.71.035105}
\bibinfo{author}{\bibfnamefont{M.}~\bibnamefont{Cococcioni}} \bibnamefont{and} \bibinfo{author}{\bibfnamefont{S.}~\bibnamefont{de~Gironcoli}}, \bibinfo{journal}{Phys. Rev. B} \textbf{\bibinfo{volume}{71}}, \bibinfo{pages}{035105} (\bibinfo{year}{2005}).

\bibitem[{\citenamefont{Hummer et~al.}(2009)\citenamefont{Hummer, Harl, and Kresse}}]{PhysRevB.80.115205}
\bibinfo{author}{\bibfnamefont{K.}~\bibnamefont{Hummer}}, \bibinfo{author}{\bibfnamefont{J.}~\bibnamefont{Harl}}, \bibnamefont{and} \bibinfo{author}{\bibfnamefont{G.}~\bibnamefont{Kresse}}, \bibinfo{journal}{Phys. Rev. B} \textbf{\bibinfo{volume}{80}}, \bibinfo{pages}{115205} (\bibinfo{year}{2009}), \urlprefix\url{https://link.aps.org/doi/10.1103/PhysRevB.80.115205}.

\bibitem[{\citenamefont{Wentzcovitch}(1991)}]{Wentzcovitch-PhysRevB-44-2358-1991}
\bibinfo{author}{\bibfnamefont{R.~M.} \bibnamefont{Wentzcovitch}}, \bibinfo{journal}{Phys. Rev. B} \textbf{\bibinfo{volume}{44}}, \bibinfo{pages}{2358} (\bibinfo{year}{1991}).

\bibitem[{\citenamefont{Kim et~al.}(1991)\citenamefont{Kim, Levin, Wentzcovitch, and Auerbach}}]{PhysRevB.44.5148}
\bibinfo{author}{\bibfnamefont{J.~H.} \bibnamefont{Kim}}, \bibinfo{author}{\bibfnamefont{K.}~\bibnamefont{Levin}}, \bibinfo{author}{\bibfnamefont{R.}~\bibnamefont{Wentzcovitch}}, \bibnamefont{and} \bibinfo{author}{\bibfnamefont{A.}~\bibnamefont{Auerbach}}, \bibinfo{journal}{Phys. Rev. B} \textbf{\bibinfo{volume}{44}}, \bibinfo{pages}{5148} (\bibinfo{year}{1991}), \urlprefix\url{https://link.aps.org/doi/10.1103/PhysRevB.44.5148}.

\bibitem[{\citenamefont{Wentzcovitch et~al.}(1993)\citenamefont{Wentzcovitch, Martins, and Price}}]{PhysRevLett.70.3947}
\bibinfo{author}{\bibfnamefont{R.~M.} \bibnamefont{Wentzcovitch}}, \bibinfo{author}{\bibfnamefont{J.~L.} \bibnamefont{Martins}}, \bibnamefont{and} \bibinfo{author}{\bibfnamefont{G.~D.} \bibnamefont{Price}}, \bibinfo{journal}{Phys. Rev. Lett.} \textbf{\bibinfo{volume}{70}}, \bibinfo{pages}{3947} (\bibinfo{year}{1993}), \urlprefix\url{https://link.aps.org/doi/10.1103/PhysRevLett.70.3947}.

\bibitem[{\citenamefont{Charpentier}(2011)}]{CHARPENTIER20111}
\bibinfo{author}{\bibfnamefont{T.}~\bibnamefont{Charpentier}}, \bibinfo{journal}{Solid State Nucl. Magn. Reson.} \textbf{\bibinfo{volume}{40}}, \bibinfo{pages}{1} (\bibinfo{year}{2011}).

\bibitem[{\citenamefont{Raghavan}(1989)}]{Raghavan1989}
\bibinfo{author}{\bibfnamefont{P.}~\bibnamefont{Raghavan}}, \bibinfo{journal}{At. Data Nucl. Data Tables} \textbf{\bibinfo{volume}{42}}, \bibinfo{pages}{189} (\bibinfo{year}{1989}).

\bibitem[{\citenamefont{Butz}(1989)}]{Butz1989}
\bibinfo{author}{\bibfnamefont{T.}~\bibnamefont{Butz}}, \bibinfo{journal}{Hyperfine Interact.} \textbf{\bibinfo{volume}{52}}, \bibinfo{pages}{189} (\bibinfo{year}{1989}).

\bibitem[{\citenamefont{Schatz and Weidinger}(1996)}]{Schatz1996}
\bibinfo{author}{\bibfnamefont{G.}~\bibnamefont{Schatz}} \bibnamefont{and} \bibinfo{author}{\bibfnamefont{A.}~\bibnamefont{Weidinger}}, \emph{\bibinfo{title}{{Nuclear condensed matter physics : nuclear methods and applications}}} (\bibinfo{publisher}{John Wiley}, \bibinfo{year}{1996}), ISBN \bibinfo{isbn}{9780471954798}.

\bibitem[{\citenamefont{Lopes et~al.}(2006)\citenamefont{Lopes, Ara\'ujo, Ramasco, Amaral, Suryanarayanan, and Correia}}]{PhysRevB.73.100408}
\bibinfo{author}{\bibfnamefont{A.~M.~L.} \bibnamefont{Lopes}}, \bibinfo{author}{\bibfnamefont{J.~P.} \bibnamefont{Ara\'ujo}}, \bibinfo{author}{\bibfnamefont{J.~J.} \bibnamefont{Ramasco}}, \bibinfo{author}{\bibfnamefont{V.~S.} \bibnamefont{Amaral}}, \bibinfo{author}{\bibfnamefont{R.}~\bibnamefont{Suryanarayanan}}, \bibnamefont{and} \bibinfo{author}{\bibfnamefont{J.~G.} \bibnamefont{Correia}}, \bibinfo{journal}{Phys. Rev. B} \textbf{\bibinfo{volume}{73}}, \bibinfo{pages}{100408} (\bibinfo{year}{2006}).

\bibitem[{\citenamefont{Barradas et~al.}(1993)\citenamefont{Barradas, Rots, Melo, and Soares}}]{PhysRevB.47.8763}
\bibinfo{author}{\bibfnamefont{N.~P.} \bibnamefont{Barradas}}, \bibinfo{author}{\bibfnamefont{M.}~\bibnamefont{Rots}}, \bibinfo{author}{\bibfnamefont{A.~A.} \bibnamefont{Melo}}, \bibnamefont{and} \bibinfo{author}{\bibfnamefont{J.~C.} \bibnamefont{Soares}}, \bibinfo{journal}{Phys. Rev. B} \textbf{\bibinfo{volume}{47}}, \bibinfo{pages}{8763} (\bibinfo{year}{1993}).

\bibitem[{\citenamefont{Correia}(1992)}]{NNFITBarr}
\bibinfo{author}{\bibfnamefont{J.~G.} \bibnamefont{Correia}}, \emph{\bibinfo{title}{Nnfit the pac manual}} (\bibinfo{year}{1992}).

\bibitem[{\citenamefont{Correia}(2018)}]{NNFIT}
\bibinfo{author}{\bibfnamefont{J.~G.} \bibnamefont{Correia}}, \emph{\bibinfo{title}{Nnfit and fft upgrades 2018}} (\bibinfo{year}{2018}).

\bibitem[{\citenamefont{Rasera and Catchen}(1993)}]{doi:10.1080/00150199308008701}
\bibinfo{author}{\bibfnamefont{R.~L.} \bibnamefont{Rasera}} \bibnamefont{and} \bibinfo{author}{\bibfnamefont{G.~L.} \bibnamefont{Catchen}}, \bibinfo{journal}{Ferroelectr.} \textbf{\bibinfo{volume}{150}}, \bibinfo{pages}{151} (\bibinfo{year}{1993}).

\bibitem[{\citenamefont{Rodríguez-Carvajal}(1993)}]{Fullprof1993}
\bibinfo{author}{\bibfnamefont{J.}~\bibnamefont{Rodríguez-Carvajal}}, \bibinfo{journal}{Physica B: Condensed Matter} \textbf{\bibinfo{volume}{192}}, \bibinfo{pages}{55} (\bibinfo{year}{1993}), ISSN \bibinfo{issn}{0921-4526}, \urlprefix\url{https://www.sciencedirect.com/science/article/pii/092145269390108I}.

\bibitem[{\citenamefont{Rodriguez-Cavajal}(2001)}]{Fullprof2001}
\bibinfo{author}{\bibfnamefont{J.}~\bibnamefont{Rodriguez-Cavajal}}, \emph{\bibinfo{title}{Recent developments of the program fullprof}} (\bibinfo{year}{2001}), \urlprefix\url{https://cir.nii.ac.jp/crid/1370283693412964373}.

\bibitem[{\citenamefont{Rodriguez-Carvajal}(1990)}]{Rodriguez1990}
\bibinfo{author}{\bibfnamefont{J.}~\bibnamefont{Rodriguez-Carvajal}}, in \emph{\bibinfo{booktitle}{satellite meeting on powder diffraction of the XV congress of the IUCr}} (\bibinfo{organization}{Toulouse, France]}, \bibinfo{year}{1990}), vol. \bibinfo{volume}{127}.

\bibitem[{\citenamefont{Kubelka and Munk}(1931)}]{Kubelka}
\bibinfo{author}{\bibfnamefont{P.}~\bibnamefont{Kubelka}} \bibnamefont{and} \bibinfo{author}{\bibfnamefont{F.}~\bibnamefont{Munk}}, \bibinfo{journal}{Z. Tech. Phys} \textbf{\bibinfo{volume}{12}}, \bibinfo{pages}{593} (\bibinfo{year}{1931}), \urlprefix\url{https://cir.nii.ac.jp/crid/1574231875746678272}.

\bibitem[{\citenamefont{Makuła et~al.}(2018)\citenamefont{Makuła, Pacia, and Macyk}}]{ReflectometryMakula}
\bibinfo{author}{\bibfnamefont{P.}~\bibnamefont{Makuła}}, \bibinfo{author}{\bibfnamefont{M.}~\bibnamefont{Pacia}}, \bibnamefont{and} \bibinfo{author}{\bibfnamefont{W.}~\bibnamefont{Macyk}}, \bibinfo{journal}{The Journal of Physical Chemistry Letters} \textbf{\bibinfo{volume}{9}}, \bibinfo{pages}{6814} (\bibinfo{year}{2018}), \eprint{https://doi.org/10.1021/acs.jpclett.8b02892}, \urlprefix\url{https://doi.org/10.1021/acs.jpclett.8b02892}.

\bibitem[{\citenamefont{Haas et~al.}(2009)\citenamefont{Haas, Tran, and Blaha}}]{PhysRevB.79.085104}
\bibinfo{author}{\bibfnamefont{P.}~\bibnamefont{Haas}}, \bibinfo{author}{\bibfnamefont{F.}~\bibnamefont{Tran}}, \bibnamefont{and} \bibinfo{author}{\bibfnamefont{P.}~\bibnamefont{Blaha}}, \bibinfo{journal}{Phys. Rev. B} \textbf{\bibinfo{volume}{79}}, \bibinfo{pages}{085104} (\bibinfo{year}{2009}), \urlprefix\url{https://link.aps.org/doi/10.1103/PhysRevB.79.085104}.

\bibitem[{\citenamefont{Streltsov et~al.}(2010)\citenamefont{Streltsov, McLeod, Moewes, Redhammer, and Kurmaev}}]{PhysRevB.81.045118}
\bibinfo{author}{\bibfnamefont{S.~V.} \bibnamefont{Streltsov}}, \bibinfo{author}{\bibfnamefont{J.}~\bibnamefont{McLeod}}, \bibinfo{author}{\bibfnamefont{A.}~\bibnamefont{Moewes}}, \bibinfo{author}{\bibfnamefont{G.~J.} \bibnamefont{Redhammer}}, \bibnamefont{and} \bibinfo{author}{\bibfnamefont{E.~Z.} \bibnamefont{Kurmaev}}, \bibinfo{journal}{Phys. Rev. B} \textbf{\bibinfo{volume}{81}}, \bibinfo{pages}{045118} (\bibinfo{year}{2010}).

\bibitem[{\citenamefont{Hinuma et~al.}(2017)\citenamefont{Hinuma, Pizzi, Kumagai, Oba, and Tanaka}}]{HINUMA2017140}
\bibinfo{author}{\bibfnamefont{Y.}~\bibnamefont{Hinuma}}, \bibinfo{author}{\bibfnamefont{G.}~\bibnamefont{Pizzi}}, \bibinfo{author}{\bibfnamefont{Y.}~\bibnamefont{Kumagai}}, \bibinfo{author}{\bibfnamefont{F.}~\bibnamefont{Oba}}, \bibnamefont{and} \bibinfo{author}{\bibfnamefont{I.}~\bibnamefont{Tanaka}}, \bibinfo{journal}{Computational Materials Science} \textbf{\bibinfo{volume}{128}}, \bibinfo{pages}{140} (\bibinfo{year}{2017}), ISSN \bibinfo{issn}{0927-0256}, \urlprefix\url{https://www.sciencedirect.com/science/article/pii/S0927025616305110}.

\bibitem[{\citenamefont{Togo et~al.}(2024)\citenamefont{Togo, Shinohara, and Tanaka}}]{togo2024textttspglibsoftwarelibrarycrystal}
\bibinfo{author}{\bibfnamefont{A.}~\bibnamefont{Togo}}, \bibinfo{author}{\bibfnamefont{K.}~\bibnamefont{Shinohara}}, \bibnamefont{and} \bibinfo{author}{\bibfnamefont{I.}~\bibnamefont{Tanaka}}, \emph{\bibinfo{title}{$\texttt{Spglib}$: a software library for crystal symmetry search}} (\bibinfo{year}{2024}), \eprint{1808.01590}, \urlprefix\url{https://arxiv.org/abs/1808.01590}.

\bibitem[{LIU(2016)}]{LIU20161}
\bibinfo{journal}{Comput. Mater. Sci.} \textbf{\bibinfo{volume}{123}}, \bibinfo{pages}{1} (\bibinfo{year}{2016}).

\bibitem[{\citenamefont{Lee et~al.}(1999)\citenamefont{Lee, James, Olsen, and Hermon}}]{Lee_JElectronMater_28_766}
\bibinfo{author}{\bibfnamefont{E.~Y.} \bibnamefont{Lee}}, \bibinfo{author}{\bibfnamefont{R.~B.} \bibnamefont{James}}, \bibinfo{author}{\bibfnamefont{R.}~\bibnamefont{Olsen}}, \bibnamefont{and} \bibinfo{author}{\bibfnamefont{H.}~\bibnamefont{Hermon}}, \bibinfo{journal}{J. Electron. Mater.} \textbf{\bibinfo{volume}{28}}, \bibinfo{pages}{766} (\bibinfo{year}{1999}).

\bibitem[{\citenamefont{Zhang et~al.}(2019)\citenamefont{Zhang, Deng, Gao, Chen, Au, Li, Yin, and Cai}}]{C9CY00997C}
\bibinfo{author}{\bibfnamefont{J.-R.} \bibnamefont{Zhang}}, \bibinfo{author}{\bibfnamefont{X.-Z.} \bibnamefont{Deng}}, \bibinfo{author}{\bibfnamefont{B.}~\bibnamefont{Gao}}, \bibinfo{author}{\bibfnamefont{L.}~\bibnamefont{Chen}}, \bibinfo{author}{\bibfnamefont{C.-T.} \bibnamefont{Au}}, \bibinfo{author}{\bibfnamefont{K.}~\bibnamefont{Li}}, \bibinfo{author}{\bibfnamefont{S.-F.} \bibnamefont{Yin}}, \bibnamefont{and} \bibinfo{author}{\bibfnamefont{M.-Q.} \bibnamefont{Cai}}, \bibinfo{journal}{Catal. Sci. Technol.} \textbf{\bibinfo{volume}{9}}, \bibinfo{pages}{4659} (\bibinfo{year}{2019}), \urlprefix\url{http://dx.doi.org/10.1039/C9CY00997C}.

\bibitem[{\citenamefont{Bayer}(1951)}]{Bayer1951}
\bibinfo{author}{\bibfnamefont{H.}~\bibnamefont{Bayer}}, \bibinfo{journal}{Z. Phys.} \textbf{\bibinfo{volume}{130}}, \bibinfo{pages}{227} (\bibinfo{year}{1951}).

\bibitem[{\citenamefont{Kushida}(1955)}]{kushida1955influence}
\bibinfo{author}{\bibfnamefont{T.}~\bibnamefont{Kushida}}, \bibinfo{journal}{J. Sci. Hiroshima Univ., Ser. A} \textbf{\bibinfo{volume}{19}}, \bibinfo{pages}{327} (\bibinfo{year}{1955}).

\bibitem[{\citenamefont{Kushida et~al.}(1956)\citenamefont{Kushida, Benedek, and Bloembergen}}]{kushida1956dependence}
\bibinfo{author}{\bibfnamefont{T.}~\bibnamefont{Kushida}}, \bibinfo{author}{\bibfnamefont{G.}~\bibnamefont{Benedek}}, \bibnamefont{and} \bibinfo{author}{\bibfnamefont{N.}~\bibnamefont{Bloembergen}}, \bibinfo{journal}{Phys. Rev.} \textbf{\bibinfo{volume}{104}}, \bibinfo{pages}{1364} (\bibinfo{year}{1956}).

\bibitem[{\citenamefont{Rocha-Rodrigues et~al.}(2020{\natexlab{a}})\citenamefont{Rocha-Rodrigues, Santos, Oliveira, Leal, Miranda, dos Santos, Correia, Assali, Petrilli, Ara\'ujo et~al.}}]{PhysRevB.102.104115}
\bibinfo{author}{\bibfnamefont{P.}~\bibnamefont{Rocha-Rodrigues}}, \bibinfo{author}{\bibfnamefont{S.~S.~M.} \bibnamefont{Santos}}, \bibinfo{author}{\bibfnamefont{G.~m. c. N.~P.} \bibnamefont{Oliveira}}, \bibinfo{author}{\bibfnamefont{T.}~\bibnamefont{Leal}}, \bibinfo{author}{\bibfnamefont{I.~P.} \bibnamefont{Miranda}}, \bibinfo{author}{\bibfnamefont{A.~M.} \bibnamefont{dos Santos}}, \bibinfo{author}{\bibfnamefont{J.~a.~G.} \bibnamefont{Correia}}, \bibinfo{author}{\bibfnamefont{L.~V.~C.} \bibnamefont{Assali}}, \bibinfo{author}{\bibfnamefont{H.~M.} \bibnamefont{Petrilli}}, \bibinfo{author}{\bibfnamefont{J.~a.~P.} \bibnamefont{Ara\'ujo}}, \bibnamefont{et~al.}, \bibinfo{journal}{Phys. Rev. B} \textbf{\bibinfo{volume}{102}}, \bibinfo{pages}{104115} (\bibinfo{year}{2020}{\natexlab{a}}), \urlprefix\url{https://link.aps.org/doi/10.1103/PhysRevB.102.104115}.

\bibitem[{\citenamefont{Rocha-Rodrigues et~al.}(2020)\citenamefont{Rocha-Rodrigues, Santos, Miranda, Oliveira, Correia, Assali, Petrilli, Ara\'ujo, and Lopes}}]{PhysRevB.101.064103}
\bibinfo{author}{\bibfnamefont{P.}~\bibnamefont{Rocha-Rodrigues}}, \bibinfo{author}{\bibfnamefont{S.~S.~M.} \bibnamefont{Santos}}, \bibinfo{author}{\bibfnamefont{I.~P.} \bibnamefont{Miranda}}, \bibinfo{author}{\bibfnamefont{G.~N.~P.} \bibnamefont{Oliveira}}, \bibinfo{author}{\bibfnamefont{J.~G.} \bibnamefont{Correia}}, \bibinfo{author}{\bibfnamefont{L.~V.~C.} \bibnamefont{Assali}}, \bibinfo{author}{\bibfnamefont{H.~M.} \bibnamefont{Petrilli}}, \bibinfo{author}{\bibfnamefont{J.~P.} \bibnamefont{Ara\'ujo}}, \bibnamefont{and} \bibinfo{author}{\bibfnamefont{A.~M.~L.} \bibnamefont{Lopes}}, \bibinfo{journal}{Phys. Rev. B} \textbf{\bibinfo{volume}{101}}, \bibinfo{pages}{064103} (\bibinfo{year}{2020}), \urlprefix\url{https://link.aps.org/doi/10.1103/PhysRevB.101.064103}.

\end{thebibliography}
\end{document}